\documentclass[journal,onecolumn]{IEEEtran}

\usepackage{amsmath, amssymb}
\usepackage{graphicx}
\usepackage{wrapfig}

\usepackage{tikz}
\usetikzlibrary{automata,positioning,decorations.pathreplacing,fit,calc}

\newcommand{\dist}{\mathrm{dist}}
\newcommand{\poly}{\mathrm{poly}}
\newcommand{\prob}{\mathrm{prob}}
\newcommand{\hash}{\mathrm{hash}}

\newcommand{\E}{\mathbb{E}}
\newcommand{\var}{\mathrm{var}}

\newtheorem{theorem}{Theorem}
\newtheorem{proposition}{Proposition}
\newtheorem{lemma}{Lemma}
\newtheorem{definition}{Definition}
\newtheorem{corollary}{Corollary}

\begin{document}

\title{On the Combinatorial Version of the Slepian--Wolf Problem%
\thanks{%
A preliminary version of this paper (excluding Theorem~2 and Section~3) was presented at the MFCS~2015.
This work was supported in part by ANR under Grant RaCAF ANR-15-CE40-0016-01.}
}

\author{%
Daniyar Chumbalov\thanks{%
Daniyar Chumbalov is with Network Dynamics Group, Ecole Polytechnique Federale de Lausanne (e-mail:~daniyar.chumbalov\symbol{64}epfl.ch). Part of this research was performed when the author was at the Moscow Institute of Physics and Technology.
}
and 
Andrei Romashchenko\thanks{%
Andrei Romashchenko is with Laboratoire d'Informatique, de Robotique et de Microélectronique de Montpellier, Centre National de la Recherche Scientifique, University of Montpellier,  on leave from IITP RAS,  Moscow (e-mail: andrei.romashchenko@lirmm.fr).
}
}

\date{}

\maketitle

\thispagestyle{plain}
\pagestyle{plain}

\begin{abstract}
\noindent
We study the following combinatorial version of the Slepian--Wolf coding scheme. Two isolated Senders are given binary strings $X$ and $Y$ respectively; the length of each string is equal to $n$, and  the Hamming distance between the strings is at most $\alpha n$. The Senders compress their strings and communicate the results to the Receiver. Then the Receiver must reconstruct both strings $X$ and $Y$. The aim is to minimize the lengths of the transmitted messages.

For an asymmetric variant of this problem (where one of the Senders transmits the input string to the Receiver without compression) with deterministic encoding a nontrivial bound was found by A.~Orlitsky and K.~Viswanathany, \cite{orlitsky}.  In our paper we prove a new lower bound for the schemes with  syndrome coding, where at least one of the Senders  uses linear encoding of the input string.

For the combinatorial Slepian--Wolf problem with randomized encoding the theoretical optimum of communication complexity was found in \cite{chumbalov}, though effective protocols with  optimal lengths of messages remained unknown. We close this gap and present  a polynomial time randomized  protocol  that achieves the optimal communication complexity. 
\end{abstract}

\section{Introduction}

\begin{figure}[h]
\centering
\includegraphics[scale=0.65]{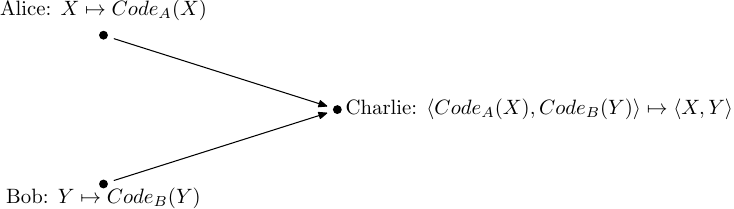}
\label{fig-scheme}
\caption{Slepian--Wolf coding scheme}\label{fig-1}
\end{figure}

\begin{figure}[h]
\centering
\includegraphics[width=0.47\textwidth]{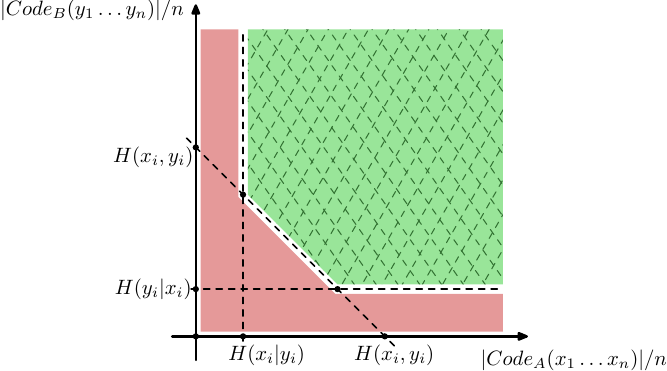}
\caption{The areas of achievable and non-achievable rates in the standard Slepian--Wolf theorem.}
\label{fig-2}
\end{figure}

The classic Slepian--Wolf coding theorem characterizes the  optimal rates for the lossless compression of two correlated data sources. In  this theorem  the correlated data sources (two sequences of correlated random variables) are encoded separately; then the compressed data are delivered to the receiver where all the data are jointly decoded, see the scheme in Fig.~\ref{fig-1}. We denote the block codes used by the Senders (Alice and Bob) as $Code_A$ and $Code_B$ respectively; the block lengths  are denoted as $|Code_A(X)|$ and $|Code_B(Y)|$ respectively (these numbers are binary logarithms of the number of codewords in the block codes for messages $X$ and $Y$  of a given length).
The seminal paper~\cite{sw} gives a very precise characterization of the profile of accessible compression rates in terms of Shannon's entropies of the sources. Namely, if the data sources are obtained as $X=(x_1\ldots x_n)$ and $Y=(y_1\ldots y_n)$, where $(x_i,y_i)$, $i=1,\ldots,n$ is a sequence of i.i.d. random pairs, then all pairs of rates satisfying the inequalities
$$
\left\{
\begin{array}{r}
|Code_A(X)| + |Code_B(Y)|\ge  H(X,Y) + o(n),\\
|Code_A(X)| \ge   H(X | Y) + o(n),\\
|Code_B(Y)| \ge  H(Y | X) + o(n),
\end{array}
\right.
$$
can be achieved (with a negligible error probability); conversely, if at least one of the inequalities 
$$
\left\{
\begin{array}{r}
|Code_A(X) + |Code_B(Y)| \ge  H(X,Y) - o(n),\\
|Code_A(X)|  \ge  H(X | Y) - o(n),\\
|Code_B(Y)|  \ge  H(Y | X) - o(n),
\end{array}
\right.
$$
is violated, then the  error probability becomes overwhelming. The areas of \emph{achievable} and \emph{non-achievable} rates are shown in Fig.~\ref{fig-2} (the hatched green area consists of achievable points, and the solid red area consists of non-achievable points; the gap between these areas vanishes as $n\to \infty$). 

It is instructive to view the Slepian--Wolf coding problem in the general context of information theory. In the  paper ``\emph{Three approaches to the quantitative definition of information}'', \cite{kolm65}, Kolmogorov  compared a \emph{combinatorial} (\emph{cf}. Hartley's combinatorial definition of information, \cite{hartley28}), a \emph{probabilistic} (\emph{cf}. Shannon's entropy), and an \emph{algorithmic} approach  (\emph{cf}. Algorithmic complexity a.k.a. Kolmogorov complexity).  Quite a few fundamental concepts and constructions in information theory have parallel implementations in all three approaches.
A prominent example of this parallelism is provided by the formal information inequalities: they can be equivalently  represented as  linear inequalities for Shannon's entropy, for Kolmogorov complexity, \cite{hrsv},  or for (logs of)  cardinalities of finite sets, \cite{rsv}, \cite{chan}. 
It is remarkable that many results known in one of these approaches look very similar to its homologues from the two other approaches, whereas the mathematical techniques and formal proofs behind them are fairly different.

As for the multi-source coding theory, two homologue theorems are known:  the  Slepian--Wolf coding theorem in Shannon's framework (where the data sources are random variables, and the achievable rates are characterized in terms of the Shannon entropies of the sources)  and  Muchnik's theorem on conditional coding, \cite{muchnik}, in Kol\-mo\-go\-rov's framework (where the data sources are words, and the achievable rates are characterized in terms of the Kolmogorov complexities of the sources). What is missing in this picture  is  a satisfactory ``combinatorial'' version of the Slepian--Wolf theorem (though several partial results are known, see blow). We try to fill this  gap; we start with  a formal definition of  the combinatorial Slepian--Wolf coding scheme  and then prove some bounds for the areas of achievable rates\footnote{I.~Csiszar  and J.~K\"orner described  the Slepian--Wolf theorem as  ``\emph{the visible part of the iceberg}'' of the multi-source coding theory; since the seminal paper by Slepian and Wolf, many parts of this ``iceberg''  were revealed and investigated, see a survey in \cite{book-korner-csisar}. Similarly,  Muchnik's theorem  has motivated numerous generalizations  and extensions in the theory of Kolmogorov complexity.  Apparently,  a similar (probably even bigger) ``iceberg'' should also exist in the combinatorial version of information theory.  However, before we explore this iceberg,  we should understand  the most basic multi-source  coding models, and  a natural starting point  is the combinatorial version of the  Slepian--Wolf coding scheme.}.

We focus on the  binary symmetric case of the  problem. In our (combinatorial) version of  the  Slepian--Wolf coding problem the data sources are binary strings, and the correlation between sources means that the Hamming distance between these strings is bounded.
More formally, we consider a communication scheme with two senders (let us call them Alice and Bob) and one receiver (we call him Charlie). We assume Alice is given a string $X$ and Bob is given a string $Y$. Both strings are of length $n$, and the Hamming distance between $X$ and $Y$ is not greater than a threshold $\alpha n$. The senders  prepare some messages $Code_A(X)$ and $Code_B(Y)$ for the receiver (i.e., Alice computes her message given $X$ and Bob computes his message given $Y$). When both messages are delivered to Charlie, he should decode them and reconstruct both strings $X$ and $Y$. Our aim is to characterize the optimal lengths of Alice's and Bob's messages.

This is the general scheme of the combinatorial version of the Slepian--Wolf coding problem.
Let us place emphasis on the most important points of our setting:
\begin{itemize}
 \item Alice knows $X$ but not $Y$ and Bob knows $Y$ but not $X$;
 \item one way communication: Alice and Bob send messages to Charlie without feedback;
 \item no communications between Alice and Bob;
 \item parameters $n$ and $\alpha$ are known to all three parties.
\end{itemize}
In some sense, this  is the ``worst case'' counterpart of the classic ``average case'' Slepian-Wolf problem.

It is usual for the theory of communication complexity to consider two types of protocols: deterministic communication protocols (Alice's and Bob's messages are deterministic functions of $X$ and $Y$ respectively, as well as Charlie's decoding function) and randomized communication protocol (encoding and decoding procedures are randomized, and for each pair $(X,Y)$ Charlie must get the right answer with only a small probability of error $\varepsilon$). 
In the next section we give the formal definitions of the deterministic and the randomized versions of the combinatorial Slepian--Wolf scheme and discuss the known lower and upper bounds for the achievable lengths of messages.

\section{Formalizing the combinatorial version of the Slepian--Wolf coding scheme}\label{sec-2}

In the usual terms of the theory of communication complexity, we study one-round communication protocols for three parties; two of them (Alice and Bob) send their messages, and the third one (Charlie) receives the messages and computes the final result. Thus, a formal definition of the communication protocol involves coding functions for Alice and Bob and the decoding function for Charlie. We are interested not only in the total communication complexity (the sum of the lengths of Alice's and Bob's messages) but also in the trade-off between the two sent messages. In what follows we formally define 
two version of the Slepian--Wolf communication scheme~--- the deterministic and the probabilistic ones.

\subsection{Deterministic communication schemes}

In the deterministic framework the \emph{communication protocol} for the combinatorial Slepian--Wolf coding scheme can be defined  simply as a pair of uniquely decodable mappings --- the coding functions of Alice and Bob.
\begin{definition}\label{unique-coding-mappings}
We say that a pair of coding mappings
\begin{equation}%
\begin{array}{rcl}
 Code_A \colon \{0,1\}^n & \to& \{0,1\}^{m_A}\\
 Code_B \colon \{0,1\}^n & \to& \{0,1\}^{m_B}
\end{array}
\nonumber
\end{equation}
is \emph{uniquely decodable}  for the combinatorial Slepian--Wolf coding scheme with parameters $(n,\alpha)$, if for each pair of images
$c_A\in \{0,1\}^{m_A}$, $c_B\in \{0,1\}^{m_B}$ there exist at most one pairs of strings $(x,y)$ such that $\dist(x,y)\le \alpha n$, and
$$
\left\{\begin{array}{rcl}
Code_A(x) &=& c_A,\\
Code_B(y) &=& c_B
\end{array}
\right.
$$
(this means that the pair $(X,Y)$ can be uniquely reconstructed given the values of $Code_A(X)$ and $Code_B(Y)$).  
If such a pair of coding mappings exists, we say that the pair of integers $(m_A,m_B)$ (the lengths of the codes) is a pair of \emph{achievable rates}.
\end{definition}

If we are interested in effective constructions of the communication scheme, we can also explicitly introduce the decoding function for Charlie  
 $$
 Decode\colon  (Code_A(X),Code_B(Y)) \mapsto (X,Y)
 $$
and investigate the computational complexities of these three mappings $Code_A$, $Code_B$ and $Decode$.

We say that \emph{encoding} in this scheme is \emph{linear} (syndrome-coding), if both functions $Code_A$ and $Code_B$ in Definition~\ref{unique-coding-mappings} can be understood as  linear mappings over the field of $2$ elements:
\begin{equation}
\nonumber
\begin{array}{rcl}
 Code_A \colon \mathbb{F}_2^n\ & \to& \mathbb{F}_2^{m_A},\\
 Code_B \colon \mathbb{F}_2^n\ & \to& \mathbb{F}_2^{m_B}.\\
\end{array}
\end{equation}
Further, we say that  an encoding is \emph{semi-linear}, if at least one of these two coding functions is linear.

\subsection{Probabilistic communication schemes}

We use the following standard\label{standard-model}
\emph{communication model with private sources of randomness}:
\begin{itemize}
\item each party (Alice, Bob, and Charlie) has her/his own ``random coin'' --- a source of random bits ($r_A$, $r_B$, and $r_C$ respectively),
\item the coins are fair, i.e.,  produce independent and uniformly distributed random bits,
\item the sources of randomness are private: each party can access only its own random coin.
\end{itemize}
In this model the message sent by Alice is a function of her input and her private random bits. Similarly, the message sent by Bob is a function of his input and his private random bits. Charlie reconstructs $X$ and $Y$ given both these messages and, if needed,  his own private random bits. (In fact, in the  protocols we construct in this paper Charlie will not use his own private random bits. The same time, the proven lower bounds remain true for protocols where Charlie employs randomness.) Let us give a more formal definition.

\begin{definition} A randomized protocol for the combinatorial Slepian--Wolf scheme with parameters $(n,\alpha,\varepsilon)$ is a triple of mappings
\begin{equation}\nonumber
\begin{array}{rcl}
 Code_A \colon \{0,1\}^n \times \{0,1\}^{R}  &  \to& \{0,1\}^{m_A},\\ \nonumber
 Code_B \colon \{0,1\}^n  \times \{0,1\}^{R}  &   \to& \{0,1\}^{m_B}, \nonumber
\end{array}
\end{equation}
and
\begin{equation}\nonumber
 Decode \colon    \{0,1\}^{m_A+m_B}   \times \{0,1\}^{R}   \to \{0,1\}^n \times \{0,1\}^n
\end{equation}
such that for every pair of strings $(X,Y)$ satisfying $\dist(X,Y)\le \alpha n$
probability (over the choice of  $r_A,r_B,r_C$) of the event 
\begin{equation}\label{def-prob-protocol}
 Decode  (Code_A(X,r_A),  Code_B(Y,r_b), r_c)=(X,Y)  
\end{equation}
 is grater than $1-\varepsilon$.
Here $m_A$ is the length of Alice's message and $m_B$ is   the length of Bob's message. The second argument of the mappings $Code_A$, $Code_B$, and $Decode$ should be understood as a sequence of random bits; we assume that each party of the protocol uses at most $R$ random bits (for some integer $R$). Condition~(\ref{def-prob-protocol}) means that 
for each pair of inputs $(X,Y) \in \{0,1\}^n\times \{0,1\}^n$ satisfying $\dist(X,Y)\le \alpha n$, the probability of the error is less than $\varepsilon$. 
\end{definition}

When we discuss \emph{efficient} communication protocols, we assume that  the mappings $Code_A$, $Code_B$, and $Decode$ can be computed in time polynomial in $n$ (in particular, this means that only $\poly(n)$ random bits can be used in the computation).

There is a major difference between the classic probabilistic setting of the Slepian--Wolf coding and the randomized protocols for combinatorial version of this problem. In the probabilistic setting we minimize the \emph{average} communication complexity (for \emph{typical} pairs $(X,Y)$);  and in the combinatorial version of the problem we deal with the \emph{worst} case communication complexity (the protocol must succeed with high probability  for \emph{each} pair $(X,Y)$ with bounded Hamming distance).

\subsection{The main results}

A simple counting argument gives very natural lower bounds for lengths of messages in the deterministic setting of the problem:
\begin{theorem}[\cite{chumbalov}]\label{trivial-bound}
For all $0<\alpha < 1/2$,
a pair  $(m_A, m_B)$ can be an achievable pair of rates for the deterministic 
combinatorial Slepian--Wolf problem with parameters $(n,\alpha)$ 
\emph{only if} the following three inequalities are satisfied
 \begin{itemize}
  \item $m_A + m_B \ge (1+h(\alpha)) n - o(n)$,
  \item $m_A \ge h(\alpha) n - o(n)$,
  \item $m_B \ge h(\alpha) n - o(n)$,
\end{itemize}  
where
$h(\alpha)$ denotes Shannon's entropy function,
$$
h(\alpha):= - \alpha \log \alpha - (1-\alpha) \log (1-\alpha).
$$
\end{theorem}
\smallskip
\noindent
 \emph{Remark 1:}  The proof of Theorem~\ref{trivial-bound} is  a straightforward counting argument. Let us observe  that the bound $m_A + m_B \ge (1+h(\alpha)) n - o(n)$ (the lower bound for the total communication complexity)
 remains valid also in the model where Alice and Bob can communicate with each other, and there is a feedback from Charlie to Alice and Bob (even if we do not count the bits sent between Alice and Bob and the bits of the feedback from Charlie).
 \begin{figure}
\center
\includegraphics[width=0.45\textwidth]{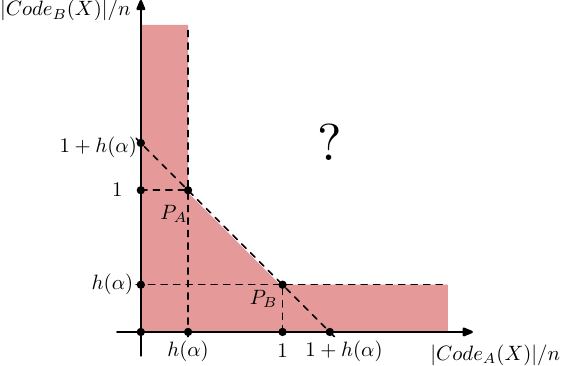}
\caption{The area of non-achievable rates.}
\label{fig-3}
\end{figure}

 \begin{figure}
\center
\includegraphics[width=0.45\textwidth]{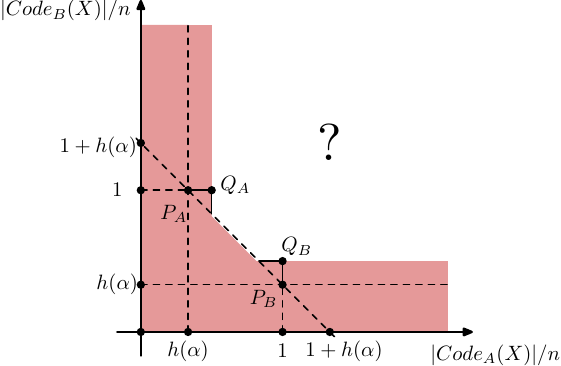}
\caption{Non-achievable rates for deterministic encoding.}
\label{fig-4}
\end{figure}
The asymptotic version of these conditions is shown in Fig.~\ref{fig-3}: the points in the  area below the dashed lines are \emph{not achievable}. 
%
Notice that these bounds are similar to the classic Slepian--Wolf bounds, see Fig.~\ref{fig-2}. The correspondence is quite straightforward:  in Theorem~\ref{trivial-bound} the sum of lengths of two messages is lower-bounded by the ``combinatorial entropy of the pair'' $(1+h(\alpha)) n$, which is basically the logarithm of the number of possible pairs $(X,Y)$ with the given Hamming distance; in the classic Slepian--Wolf theorem the sum of two channel capacities is bounded by the Shannon entropy of the pair. Similarly, in Theorem~\ref{trivial-bound} the lengths of both messages are bounded  by $h(\alpha) n$, which is  the ``combinatorial conditional entropy'' of $X$ conditional on  $Y$ or $Y$ conditional on $X$, i.e., the logarithm of the maximal number of $X$'s compatible with a fixed $Y$ and vice-versa; in the standard Slepian--Wolf theorem the corresponding quantities are bounded by the two conditional Shannon entropies.

 Though the trivial bound from Theorem~\ref{trivial-bound} looks very similar to the lower bounds in the classic Slepian--Wolf theorem and in Muchnik's conditional coding theorem, this parallelism cannot be extended  further.  In fact, the bound from Theorem~\ref{trivial-bound} is not optimal (for the deterministic communication protocols). Actually we cannot achieve any pairs of code lengths in  $\Theta(n)$-neighborhoods of the  points $(n, h(\alpha)n)$ and $(h(\alpha)n, n)$ (around the points $P_A$ and $P_B$ in Fig.~\ref{fig-3}).
This negative result was proven by Orlitsky and Viswanathany in \cite{orlitsky}, see also a discussion in \cite{chumbalov}.
More specifically, \cite{orlitsky} analyzes an asymmetric version of the Slepian--Wolf scheme and proves a lower bound for  the length of $Code_A(X)$ assuming that $Code_B(Y)=Y$. Technically, \cite{orlitsky}  shows that for some $F(\alpha)>h(\alpha)$  the pair of rates $(F(\alpha)n,n)$ is not accessible  (i.e., the point $Q_A$ and  its symmetric counterpart $Q_B$ in Fig.~\ref{fig-4} are not achievable). 
The proof in \cite{orlitsky}  employs the techniques from coding theory (see Proposition~\ref{prop-orlitsky} in Appendix);  the value of $F(\alpha)$ can be chosen as the best known lower bound for the rate of an error correcting code that can handle the fraction of errors $\alpha$.

Though this argument deals with only very special type of schemes where $Code_B(Y)=Y$, it  also implies some bound for the general Slepian--Wolf problem. Indeed, since the points $Q_A$ and  $Q_B$ are not achievable, we can conclude that all points downwards and to the left from these points are  non achievable either (by decreasing the rates we make the communication problem only harder). So we can exclude two  right triangles with  a vertical and horizontal legs meeting at points $Q_A$ and $Q_B$, see Fig.~\ref{fig-4}. Further, if some point $(m_A,n)$ is not achievable, than all points $(m_A,n')$ with $n'>n$ cannot  be achievable either (at the rate $n$ Bob can communicate the entire value of $Y$, so increasing the capacity of Bob's channel cannot help any more). Hence, we can exclude the entire vertical stripe to the left from $Q_A$ and symmetrically the entire horizontal stripe below $Q_B$, as shown in Fig.~\ref{fig-4}.
Thus, the bound from Theorem~\ref{trivial-bound} does not provide the exact characterization of the set of achievable pairs. Here we see a sharp contrast with the classic  Slepian--Wolf coding.

In this paper we prove another negative result for all \emph{linear} and even for all \emph{semi-linear} encodings:
\begin{theorem}\label{thm-linear-codes-bound}
For every $\alpha\in(0,\frac14)$   there exists  an $\alpha'\in(\alpha,\frac12)$ with the following property. 
For all  pairs $(m_A,m_B)$ achievable for  the semi-linear deterministic combinatorial Slepian--Wolf scheme with the distance $\alpha$,  it holds that
\begin{eqnarray}\label{eq-thm-linear-codes-bound}
m_A+m_B \ge (1+h(\alpha'))n-o(n).
\end{eqnarray}
Moreover, the value of $\alpha'$ can be defined explicitly as
\begin{eqnarray}\label{eq-johnson}
\alpha' := \frac{1-\sqrt{1-4\alpha}}{2}.
\end{eqnarray}
\end{theorem}
\begin{figure}
\centering
\includegraphics[width=0.45\textwidth]{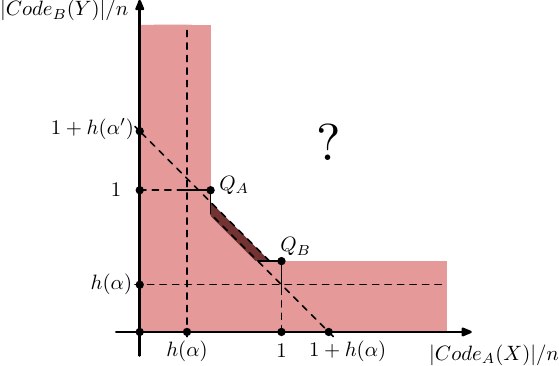}
\caption{Non-achievable rates for deterministic linear   encoding.}
\label{fig-5}
\end{figure}

The geometrical meaning of Theorem~\ref{thm-linear-codes-bound} is shown in Fig.~\ref{fig-5}:  for every $\alpha<1/4$ 
all pairs of rates below the line $m_A+m_B = (1+h(\alpha'))n-o(n)$ are not achievable. This is a strictly better bound than 
the condition $m_A+m_B \ge (1+h(\alpha))n-o(n)$ from Theorem~\ref{trivial-bound}. Notice that around the points $Q_A$ and $Q_B$
the bound from \cite{orlitsky}  remains stronger than \eqref{eq-thm-linear-codes-bound} (we borrowed the light red area from Fig.~\ref{fig-4}). This happens because 
the McEliece--Rodemich--Rumsey--Welch bound
(plugged in the proof of the bound in \cite{orlitsky}) is stronger than the Elias--Bassalygo bound (implicitly used in the proof of Theorem~\ref{thm-linear-codes-bound}).
The area where \eqref{eq-thm-linear-codes-bound} is better than any other known bound is shown  in Fig.~\ref{fig-5} in dark red color.

It is instructive to compare the known necessary and sufficient conditions for the achievable rates. If we plug  some linear codes approaching the Gilbert--Varshamov bound in the construction from \cite[theorem~2]{chumbalov}, we obtain the following proposition.
\begin{proposition}\label{prop-chumbalov}
For each real $\alpha\in(0,\frac14)$ there exists a function $\delta(n)=o(n)$ such that 
and for all $n$, all pairs of integers $(m_A,m_B)$ satisfying
\begin{itemize}
\item $m_A+m_B \ge (1+h(2\alpha))n+\delta(n)$,
\item $m_A \ge h(2\alpha)n+\delta(n)$,
\item $m_B \ge h(2\alpha)n+\delta(n)$
\end{itemize}
are achievable for the deterministic combinatorial Slepian--Wolf scheme with parameters $(n,\alpha)$. Moreover, these rates can be achieved with some \emph{linear} schemes (where encodings of Alice and Bob are linear).
\end{proposition}

\begin{figure}[ht]
\centering
\includegraphics[width=0.45\textwidth]{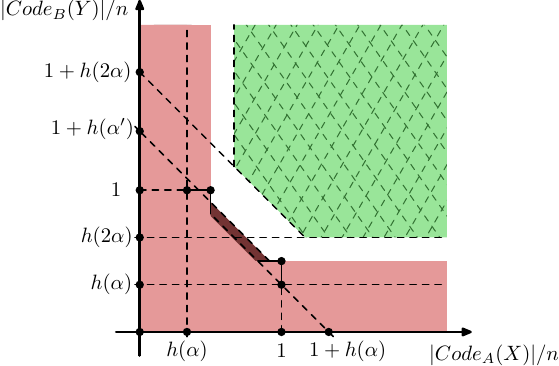}
\caption{Achievable and non-achievable rates for deterministic linear  encoding.}
\label{fig-6}
\end{figure}

In Fig.~\ref{fig-6} we combine together the known upper and lower bounds: the points in the light red area are  non-achievable (for any deterministic scheme) due to Theorem~\ref{trivial-bound} and  \cite{orlitsky}; the points in the dark red area are non-achievable (for linear and semi-linear deterministic scheme) by Theorem~\ref{thm-linear-codes-bound}; the points in the hatched green area are achievable due to Proposition~\ref{prop-chumbalov}. The gap between the known  sufficient and necessary conditions remains pretty  large.

\begin{figure}[ht]
\centering
\includegraphics[width=0.45\textwidth]{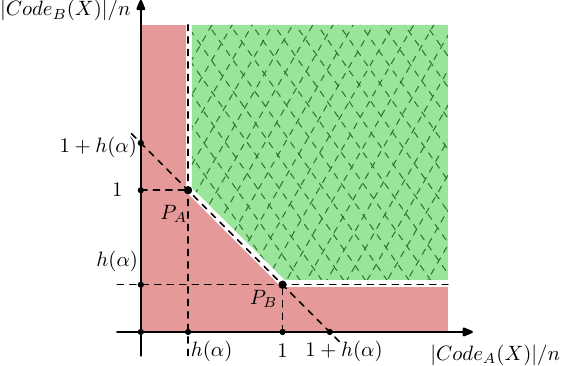}
\label{fig-sw-areas}
\caption{The areas of achievable and non-achievable rates for randomized protocols.}
\label{fig-7}
\end{figure}

Our proof of Theorem~\ref{thm-linear-codes-bound} (see Section~\ref{lower-bound}) is inspired by the classic proof of the Elias--Bassalygo bound from coding theory, \cite{bassalygo}. Our usage of the Elias--Bassalygo bound 
is not black-box: we use the proofs rather than the statement of these theorems. This explains why we cannot  employ instead of  the Elias--Bassalygo bound   any other bound from coding theory,  and therefore there is no simple way to improve \eqref{eq-johnson}. Also, we do not know whether this bound holds for non-linear encodings.  This seems to be an interesting question in between coding theory and communication complexity theory. 

Thus, we see that the solution of the deterministic version of the combinatorial Slepian--Wolf  problem of is quite different from the standard Slepian--Wolf  theorem. What about the probabilistic version?
The same conditions as in Theorem~\ref{thm-chumbalov} hold for the probabilistic protocols:
\begin{theorem}[\cite{chumbalov}]\label{thm-chumbalov}
For all  $\varepsilon\ge 0$ and $0<\alpha < 1/2$,
a pair  $(m_A, m_B)$ can be an achievable pair of rates for the probabilistic 
combinatorial Slepian--Wolf problem with parameters $(n,\alpha,\varepsilon)$ 
\emph{only if} the following three inequalities are satisfied
 \begin{itemize}
  \item $m_A + m_B \ge (1+h(\alpha)) n - o(n)$,
  \item $m_A \ge h(\alpha) n - o(n)$,
  \item $m_B \ge h(\alpha) n - o(n)$.
\end{itemize}
\end{theorem}
\smallskip
\noindent
 \emph{Remark 2:}  The bounds from Theorem~\ref{thm-chumbalov} holds also for the model with public randomness, where all the parties access a common source of random bits.
 
In the contrast to the deterministic case,  for the probabilistic setting the sufficient conditions for achievable pairs are very close to the basic lower bound above. More precisely, for every $\varepsilon>0$, all pairs in the hatched (green) area in Fig.~\ref{fig-7} are achievable for  the combinatorial Slepian--Wolf problem with parameters $(n,\alpha,\varepsilon)$, see \cite{chumbalov}. The gap between known necessary and sufficient conditions (the hatched and non-hatched areas in the figure) vanishes as $n$ tends to infinity.  Thus, for randomized  protocols we get a result similar to the classic Slepian--Wolf  theorem. 

So, the case of randomized protocol for the combinatorial Slepian--Wolf problem seems closed: the upper and lower bounds known from \cite{chumbalov} (asymptotically) match each other.
The only annoying shortcoming of the result in \cite{chumbalov} was computational complexity. The  protocols in \cite{chumbalov} require exponential computations on the senders and  the receiver sides. In this paper we improve computational complexity of these protocols without  degrading communication complexity. We propose a communication protocol with (i) optimal trade-off between the lengths of senders messages and (ii) polynomial time algorithms for all parties. More precisely, we prove the following theorem\footnote{Not surprisingly, the inequalities in Theorem~\ref{thm-chumbalov} and Theorem~\ref{thm-main} are very similar. The gap between necessary and sufficient conditions for achievable pairs is only $o(n)$.}: 

\begin{theorem}
\label{thm-main}
There exists a real $d>0$ and a function $\delta(n)=o(n)$ such that for all  $0<\alpha < 1/2$ and all integers $n$, 
every pair  $(m_A, m_B)$ that satisfies  three inequalities 
 \begin{itemize}
   \item $m_A + m_B \ge (1+h(\alpha)) n + \delta(n)$,
  \item $m_A \ge h(\alpha) n + \delta(n)$,
  \item $m_B \ge h(\alpha) n + \delta(n)$,
 \end{itemize}
 is achievable  for the combinatorial Slepian--Wolf coding problem with parameters 
$(n, \alpha,$ $\varepsilon(n)=2^{-\Omega(n^{d})})$ (in the  communication model with private sources of randomness). Moreover, all the computations in the communication protocol can be done in polynomial time.
\end{theorem}
Poly-time protocols achieving the marginal pairs $(n, h(\alpha)n+o(n))$ and  $(h(\alpha)n+o(n), n)$  were originally proposed in \cite{smith} and later in \cite{chuklin}.  We generalize these results: we construct effective protocols for all points in hatched  area in Fig.~\ref{fig-7}. In fact, our construction uses the techniques proposed in \cite{smith} and  \cite{smith-guruswami-2}.

By definition of the combinatorial Slepian--Wolf coding problem, our proof of Theorem~\ref{thm-main} provides a pair of randomized codes and a decoding function, which succeed with high probability on \emph{every} pair of valid inputs (i.e., on every pair of inputs with bounded Hamming distance). If we fix the random bits used in our encoding, we obtain a deterministic scheme that succeeds on \emph{most} pairs of inputs. Moreover, we can fix the values of these random bits to a simply computable sequence, so that the resulting deterministic scheme succeeds on most pairs of inputs, and the encoding and decoding procedures can be performed in  polynomial time.
Such a scheme gives a solution for the classic (probabilistic) Slepian--Wolf problem  with asymptotically optimal communication complexity, though the constants hidden in the small-o notation would be worse than in most known constructions. In particular, the resulting $o(n)$-terms would worse that those from the poly-time computable scheme in \cite{uyematsu} discussed below. However, the main challenge of our proof is to construct a scheme that work well  not for the majority but for \emph{all} pairs of inputs.

The rest of this paper is organized as follows. In section~\ref{lower-bound} we discuss a non-trivial lower bound for communication complexity of
the deterministic version of the combinatorial Slepian--Wolf coding scheme (a proof of Theorem~\ref{thm-linear-codes-bound}). The argument employs binary Johnson's bound, similarly to the proof of the well known Elias--Bassalygo bound in coding theory.

In Section~\ref{section-randomized} we provide an effective protocol for the randomized version of the combinatorial Slepian--Wolf coding scheme 
(a proof of Theorem~\ref{thm-main}).
Our argument combines several technical tools:
reduction of one global coding problem with strings of length $n$ to many local problems with strings of length $\log n$ (similar to the classic technique of concatenated codes); Reed--Solomon checksums; pseudo-random permutations; universal hashing.  
Notice that a similar  technique of concatenated codes combined with  syndrome encoding was used in \cite{uyematsu} to construct an efficient version of  the classic Slepian--Wolf scheme (which is allowed to fails on a small fraction of input pairs). Our construction is technically more involved since we need to succeed on  \emph{all} valid pairs of inputs. The price that we pay for this is a slower convergence to the asymptotical limits:  the remainder terms $o(n)$ from our proof are larger than those from \cite{uyematsu}.

In conclusion we discuss how to make the protocol from Theorem~\ref{thm-main} more practical --- how to simplify the algorithms involved in the protocol. The price for this simplification is a weaker bound for the probability of error.

\subsection{Notation}
Through this paper, we use the following notation:

\begin{itemize}
\item we denote 
 $
 h(\alpha):=\alpha\log(1/\alpha)+(1-\alpha)\log1/(1-\alpha),
 $
and use the standard  asymptotic bound for the binomial coefficients:
 $
 {n\choose {\alpha n} }= 2^{h(\alpha) n + O(\log n)},  
 $ 
\item we denote by $\omega(x)$ the \emph{weight} (number of $1$'s) in a binary string $x$,
\item for a pair of binary strings $x,y$ of the same length we denote by $x\oplus y$ their bitwise sum modulo $2$,
\item we denote by $\dist(v,w)$ the Hamming distance between  bit strings $v$ and $w$ (which coincides with $\omega(x\oplus y)$),
\item For an $n$-bits string  $X=x_1\ldots x_n$ and a tuple of indices $I=\langle i_1,\ldots,i_s\rangle$ we denote  $
 X_I := x_{i_1} \ldots x_{i_s}.
 $
\end{itemize}

\section{Lower bounds for deterministic protocols with semi-linear encoding}\label{lower-bound}

In this section we prove Theorem~\ref{thm-linear-codes-bound}.
We precede the proof of this theorem by several lemmas. First of all, we define the notion of  \emph{list decoding} for 
the  Slepian--Wolf scheme 
(similar to the standard notion of list decoding from coding theory).

\begin{definition}
We say that a pair of coding mappings
\begin{equation}\label{coding-mappings}
\begin{array}{rcl}
Code_A \colon \{0,1\}^n & \to& \{0,1\}^{m_A}\\
Code_B \colon \{0,1\}^n & \to& \{0,1\}^{m_B}
\end{array}
\end{equation}
is \emph{$L$-list decodable}  for the combinatorial Slepian--Wolf coding scheme with parameters $(n,\alpha)$, if for each pair of images
$c_A\in \{0,1\}^{m_A}$, $c_B\in \{0,1\}^{m_B}$ there exist at most $L$ pairs of strings $(x,y)$ such that $\dist(x,y)\le \alpha n$, and
$$
\left\{\begin{array}{rcl}
Code_A(x) &=& c_A,\\
Code_B(y) &=& c_B.
\end{array}
\right.
$$
\end{definition}

The lengths of codewords of $\poly(n)$-decodable mappings must obey effectively  
the same asymptotical bounds as  the codewords of uniquely decodable mappings.  Let us formulate this statement more precisely.

\begin{lemma}\label{lemma-list-decoding-lower-bound}
If $(m_A,m_B)$ is an achievable pair of integers for the combinatorial Slepian--Wolf scheme with parameters  $(n,\alpha)$ with list decoding \textup(with the list size $L=\poly(n)$\textup), then
\begin{itemize}
\item $m_A+m_B \ge (1+h(\alpha))n-o(n)$,
\item $m_A \ge h(\alpha)n-o(n)$,
\item $m_B \ge h(\alpha)n-o(n)$.
\end{itemize}
\end{lemma}
The lemma follow from a standard counting argument. 
The lower bounds in this lemma are asymptotically the same as the bounds for the schemes with unique decoding 
in Theorem~\ref{thm-chumbalov}. The difference between the right-hand side of the inequalities in this lemma and in Theorem~\ref{thm-chumbalov} 
is only $\log L$, which is negligible (an $o(n)$-term) as $L=\poly(n)$.

We will use the following well known bound from coding theory.

\begin{lemma}[Binary Johnson's bound]\label{lemma-johnson}
Let $\alpha$ and $\alpha'$ be positive reals  satisfying (\ref{eq-johnson}). Then for every list  of n-bits strings  $v_i$,
$$
v_1, \ldots, v_{2n+1} \in\{0,1\}^n
$$
with Hamming weights at most $\alpha' n$ \textup(i.e., all $v_i$ belong to the ball of radius $\alpha ' n$ around $0$ in  Hamming's metrics\textup),  
there exists a pair of strings $v_{i}$, $v_j$  \textup($i\not= j$\textup) such that
 $$
 \dist(v_i, v_j) \le 2\alpha n.
 $$ 
\end{lemma}
\emph{Comment:}
Johson's bounds were suggested in \cite{johnson-bound-old} as limits on the size of error-correcting codes. 
Several extensions and generalizations of these bounds were found in subsequent works, see  \cite{johnson-bound}.
An elementary and self-contained proof of Lemma~\ref{lemma-johnson} can be found also in \cite{johnson-bound-elementary}.

Now we are ready to prove the main technical lemma: every pair of mappings that is \emph{uniquely} decodable for the Slepian--Wolf scheme
with parameters $(n,\alpha)$ must be also $\poly(n)$-decodable  with parameters $(n,\alpha')$ with some  $\alpha'>\alpha$.
The next lemma is inspired by the proof of Elias--Bassalygo's bound.

\begin{lemma}\label{lemma-elias-bassalygo}
Let $\alpha$ and $\alpha'$ be positive reals as in (\ref{eq-johnson}).
If a pair of integers $(m_A,m_B)$ is achievable  for the combinatorial semi-linear Slepian--Wolf scheme with parameters  $(n,\alpha)$ (with unique decoding), then the same
pair  is  achievable for the combinatorial Slepian--Wolf scheme for the greater distance  $\alpha'$ with $(2n)$-list decoding.
The value of $\alpha'$ can be explicitly defined from  (\ref{eq-johnson}).
\end{lemma}
\begin{IEEEproof} Let as fix some pair of encodings
$$
\begin{array}{rcl}
Code_A \colon \{0,1\}^n & \to& \{0,1\}^{m_A},\\
Code_B \colon \{0,1\}^n & \to& \{0,1\}^{m_B}
\end{array}
$$
that is \emph{uniquely  decodable} for pairs $(x,y)$ with the Hamming distance $\alpha n$. We assume that at least one of these mappings (say, $Code_A$) is linear. 
To prove the lemma we  show that the same pair of encodings is \emph{list} $\poly(n)$-\emph{list decodable} for the pairs of strings with a  greater Hamming distance $\alpha' n$. 

Let us fix some  $c_A\in  \{0,1\}^{m_A}$ and $c_B\in  \{0,1\}^{m_B}$, and take the list of all $Code_A$- and $Code_B$-preimages of these points:
\begin{itemize}
\item let $\{x_i\}$ be all strings such that $Code_A(x_i)=c_A$, and
\item let $\{y_j\}$ be all strings such that $Code_B(y_j)=c_B$.
\end{itemize}
Our aim is to prove that the number of pairs $(x_i, y_j)$ such that $\dist(x_i, y_j)\le \alpha' n$ is not greater than $2n$. Suppose for the sake of contradiction that the number of such pairs is at least $2n+1$.

For each pair $(x_i, y_j)$ that satisfies $\dist(x_i, y_j)\le \alpha' n$ we take their bitwise sum,  $v := x_i\oplus y_j$. Since the Hamming distance between $x_i$ and $y_j$ is not greater than $\alpha' n$, the weight of  $v$ is not greater than $\alpha' n$. Thus, we get at least $2n+1$ different strings $v_s$ with Hamming weights not greater than $\alpha' n$. From Lemma~\ref{lemma-johnson} it follows that there exist a pair of strings $v_{s_1}$, $v_{s_2}$ (say, $v_{s_1} = x_{i_1}\oplus y_{j_1}$ and $v_{s_2} = x_{i_2}\oplus y_{j_2}$) such that $\dist(v_{s_1}, v_{s_2})\le 2\alpha n$. Hence, there exists a string $w$  that is $(\alpha n)$-close to both $v_{s_1}$ and $v_{s_2}$, i.e., 
 $$
 \dist(v_{s_1}, w)\le \alpha n
 \mbox{ and }
 \dist(v_{s_2}, w)\le \alpha n. 
 $$
We use this $w$ as a translation vector and define
$$
z_1 :=  x_{i_1} \oplus w
\mbox{ and }
z_2 :=  x_{i_2} \oplus w.
$$
For the chosen $w$ we have 
$$
\dist( z_1 , y_{j_1})  = \omega(x_{i_1}\oplus w\oplus y_{i_1}) = \dist(v_{s_1},w) \le \alpha n
$$
and
$$
\dist( z_2 , y_{j_2})  = \omega(x_{i_2}\oplus w\oplus y_{i_2}) = \dist(v_{s_2},w) \le \alpha n.
$$
Further, since  $Code_A(x_{i_1})=  Code_A(x_{i_2})=c_A$ and the mapping $Code_A$ is linear, we get
$Code_A(x_{i_1}\oplus w) =  Code_A(x_{i_2}\oplus w)$. Hence,
$$
\begin{array}{l}
Code_A(z_1) = Code_A(x_{i_1}\oplus w) =  Code_A(x_{i_2}\oplus w) 
=  Code_A(z_2). \nonumber
\end{array}
$$
Thus, we obtain two different pairs of strings $(z_1, y_{j_1})$ and $(z_2, y_{j_2})$ with the Hamming distances bounded by $\alpha n$, such that 
$$
\left\{
\begin{array}{lcl}
Code_A(z_1) &=& Code_A(z_2), \\
Code_B(y_{j_1}) &=& Code_B(y_{j_2}).
\end{array}
\right.
$$
This contradicts the assumption that the codes $Code_A$  and $Code_B$ are uniquely decodable for pairs at the distance $\alpha n$. The lemma is proven.
\end{IEEEproof}

\smallskip

Now we can prove Theorem~\ref{thm-linear-codes-bound}. Assume that a pair of integers $(m_A,m_B)$ is achievable for the combinatorial Slepian--Wolf coding scheme with unique decoding for a distance $\alpha n$. From Lemma~\ref{lemma-elias-bassalygo} it follows that the same pair is achievable for the combinatorial Slepian--Wolf coding scheme with $(2n)$-list decoding with a greater distance $\alpha' n$. Then, we apply the first inequality of Lemma~\ref{lemma-list-decoding-lower-bound} and get the required bound for $m_A$ and $m_B$.

\section{Randomized polynomial time protocol}
\label{section-randomized}

\subsection{Some technical tools}

In this section we summarize  the technical tools that we use to construct an effective randomized protocol.

\subsubsection{Pseudo-random permutations}

\begin{definition}
For a pair of distributions $\rho_1$, $\rho_2$ on a finite set $S$ we call by the \emph{distance} between $\rho_1$ and $\rho_2$ the sum
 $$
 \sum\limits_{x\in S} |\rho_1(x) - \rho_2(x)|
 $$
 (which is the standard $l_1$-distance, if we understand the distributions as vectors whose dimension is equal to the size of $S$).

A distribution on the set $S_n$ of permutations of $\{1,\ldots,n\}$ is called \emph{almost $t$-wise independent} if for every tuple of indices $1\le i_1 <i_2<\ldots<i_t\le n$, the distribution of $(\pi(i_1), \pi(i_2), \ldots, \pi(i_t))$ for $\pi$ chosen according to this distribution has distance at most $2^{-t}$ from the uniform distribution on $t$-tuples of $t$ distinct elements from $\{1,\ldots,n\}$.
\end{definition}

\begin{proposition}[\cite{knr05}]\label{proposition-knr}
For all $1\le t\le n$, there exists an integer $T=O(t\log n)$ and an explicit map
 $
 \Pi : \{0,1\}^T \to S_n,
 $
computable in time $\poly(n)$, such that the distribution $\Pi(s)$ for random $s\in \{0,1\}^T$
is almost $t$-wise independent. 
\end{proposition}

\subsubsection{Error correcting codes}
\begin{proposition}[Reed-Solomon codes] \label{proposition-rs}
Assume $m+2s<2^k$. Then we can assign to every sequence of $m$ strings 
  $
  X=\langle X^1, \ldots, X^m \rangle
  $ 
  (where $X^j\in\{0,1\}^k$ for each $j$)  a string  of \emph{checksums}  $Y=Y(X)$  of length $(2s+1)k$,
   $$
Y\ :\ \{0,1\}^{km} \to \{0,1\}^{(2s+1)k}
   $$
 with the following property. If at most $s$ strings $X^j$ are corrupted,  the initial tuple $X$ can be  uniquely reconstructed given the value of $Y(X)$.  Moreover, encoding (computation  $X\mapsto Y(X)$) end decoding (reconstruction of the initial values of $X$)  can be done in time $\poly(2^k)$. 
\end{proposition}
\begin{IEEEproof} 
The required construction can be obtained from  a systematic Reed--Solomon code with suitable parameters (see, e.g., \cite{coding-book}). Indeed, we can think of  $X = \langle X^1, \ldots, X^m \rangle$ as of a sequence of elements in a finite field $\mathbb{F}= \left\{ q_1, q_2, \dots, q_{2^k}\right\}$. Then, we interpolate a polynomial $P$ of degree at most $m-1$ such that $P(q_i) = X_i$ for $i=1,\ldots, m$ and take   the values of $P$ at some other points of the field as checksums:
 $$ Y(X) := \langle P(a_{m+1}), P(a_{m+2}), \dots, P(a_{m+2s+1})\rangle.$$ 
 The tuple $$\langle X^1, \ldots, X^m, P(a_{m+1}), P(a_{m+2}), \dots, P(a_{m+2s+1}) \rangle$$ is a codeword of the Reed--Solomon code, and we can recover it if at most $s$ items of the tuple are corrupted. It is well known that the error-correction procedure for Reed--Solomon codes can be implemented in polynomial time.
\end{IEEEproof}

\subsubsection{Universal hashing}
\begin{proposition}[universal hashing family, \cite{hash-func}]\label{proposition-hash}
There exists a family of poly-time computable functions
 $$
 \hash_i : \{0,1\}^n \to \{0,1\}^k
 $$
 such that  $\forall x_1,x_2\in  \{0,1\}^n$, $x_1\not= x_2$ it holds
  $$
  \prob_i[\hash_i(x_1) = \hash_i(x_2)] = 1/2^k,
  $$
where index  $i$ ranges over  $\{0,1\}^{O(n+k)}$ (i.e.,  each hash function from the family can be specified by a string of length $O(n+k)$ bits). 

Such a family of hash functions can be constructed explicitly: the value of  $\hash_i(x)$ can be computed in polynomial time from $x$ and $i$.
\end{proposition}

The parameter $k$ in Proposition~\ref{proposition-hash} is referred to as \emph{the length of the hash}.

The following claim is an (obvious) corollary of the definition  of a universal hashing family. Let $\hash_i(x)$ be a family of functions satisfying Proposition~\ref{proposition-hash}.
Then for every  $S\subset \{0,1\}^n$, for each $x\in S$,
  $$
  \prob_i[ \exists x'\in S \mbox{ s.t. } x'\not=x \mbox{ and } \hash_i(x)=\hash_i(x')] < \frac{|S|}{2^k}.
  $$
This property allows to identify an element in $S$ by its hash value.

\subsubsection{The law of large numbers for $t$-independent sequences}

The following version of the law of large numbers is suitable for our argument:
\begin{proposition}[see \cite{dhrs94,sss95,smith}]
\label{proposition-lln}
Assume $\xi_1,\ldots,\xi_m$ are random variables ranging over $\{0,1\}$, each with expectation at most $\mu$, and for some $c<1$, for every set of $t=m^c$  indices $i_1, \ldots, i_t$ we have
 $$
 \prob[\xi_{i_1}= \ldots = \xi_{i_t} =1] \le \mu^t.
 $$
If $t\ll \mu m$, then
 $$
\prob\left[\sum\limits_{i=1}^m \xi_i > 3\mu m\right]  =  2^{-\Theta(m^{c}) }.
 $$  
\end{proposition}

More technically, we will use the following lemma:

\begin{lemma} \label{lemma1} 
(a) Let $\rho$ be a positive constant,   $k(n)=\log n$, and $\delta=\delta(n)$ some function of $n$. Then
for  each pair of subsets $\Delta, I \subset  \{1, \ldots,  k \}$  such that $|\Delta|=k$  and $|I| = \rho n$, for a $k$-wise almost independent  permutation  
$\pi : \{1,\ldots, n\} \to \{1,\ldots, n\}$,
 $$
 \mu := \prob_{\pi}\big[\  \big| \ |\pi( I ) \cap \Delta|  - \rho k\  \big| > \delta k\ \big]  = O\left(\frac1{\delta^2 k}\right).
 $$
 
 (b) Let $\{1,\ldots, n \} = \Delta_1 \cup \ldots \cup \Delta_m$, where $\Delta_j$ are disjoint sets of cardinality $k$ (so $m = n/k$).
Also we let  $t=m^{c}$ (for some $c<1$) and  assume ${t} \ll \mu m$. Then,  for a $(tk)$-wise almost independent permutation 
$\pi$, the probabilities
 $$ 
 \prob_{\pi}\big[\  \big| \pi( I ) \cap \Delta_j \big| > (\rho+\delta) k \  \mbox{ for  } {}\ge 3\mu m \mbox{ different }j    \big]   
$$
and
$$
 \prob_{\pi}\big[\  \big| \pi( I ) \cap \Delta_j \big| < (\rho-\delta) k \  \mbox{ for   } {}\ge 3\mu m \mbox{ different }j    \big]   
 $$ 
are both not greater than $2^{-\Theta(m^{c})}$.
\end{lemma}
(The proof is deferred to Section~\ref{appendix}.)

Notice that a uniform distribution on the set of all permutation is a special case of a $k$-wise almost independent permutation. So the claims of Lemma~\ref{lemma1} can be applied to a uniformly chosen random permutation.

\subsection{Auxiliary communication models: shared and imperfect randomness} \label{section-poly-time}

The complete proof of Theorem~\ref{thm-main} involves a few different technical tricks. To make the construction more modular and intuitive, we split it in several possibly independent parts. To this end,  we  introduce several auxiliary communication models. The first two models are somewhat artificial; they are of  no independent interest, and  make sense only as intermediate steps of the proof of the main theorem. Here is the list of our communication model:

\smallskip
\noindent
\textbf{Model 1. The model with partially shared sources of perfect randomness:}  Alice and Bob have their own sources of independent uniformly distributed random bits. Charlie has a free access to Alice's and Bob's sources of randomness (these random bits are not included in the communication complexity); but Alice and Bob cannot access the random bits of each other. 

\smallskip
\noindent
\textbf{Model 2. The model with partially shared sources of $T$-non-perfect randomness:}  Alice and Bob have their own (independent of each other) sources of  randomness. However these sources are not perfect: they can produce   $T$-independent sequences of bits and $T$-\emph{wise almost independent} permutations on $\{1,\ldots,n\}$. Charlie has a free access to Alice's and Bob's sources of randomness, whereas Alice and Bob cannot access the random bits of each other. 

\smallskip
\noindent
\textbf{Model 3. The standard model with private sources of perfect randomness (our main model):}  In this model Alice and Bob have their own sources of independent uniformly distributed random bits. Charlie cannot access random bits of  Alice and Bob unless they include these bits in their messages. 

\smallskip

We stress that we variate only the rules of access to the auxiliary  random bits;
in all these models, Alice and Bob access their own inputs (strings $x$ and $y$) but cannot access the inputs of each other.

We show that in all these models the profile of achievable pairs of rates is the same as in Theorem~\ref{thm-chumbalov} (the hatched area in Fig.~\ref{fig-7}).
We start with an effective protocol for  Model~1, and then extend it to Model~2, and at  last to Model~3.

\subsection{An effective protocol for Model~1 (partially shared sources of perfect randomness)}\label{section-model-2}
\label{section-protocol-model-1}
In this section we show that all pairs of rates from the hatched area in ~4 are achievable for Model~1. Technically, we prove the following statement.
\begin{proposition}\label{prop-model-2}
The version of Theorem~\ref{thm-main} holds for the Communication Model~1.
\end{proposition}

\smallskip

\emph{Remark 1.} Our protocol involves random objects of different kinds:  randomly chosen permutations and random hash functions from a universal family.  In this section we assume that the used randomness is perfect. This means that  all permutations are chosen with the uniform distribution, and all hash functions are chosen independently.

\subsubsection{Parameters of the construction}\label{section-parameters}
Our construction has some ``degrees of freedom'';  it  involves several parameters, and values of these parameters can be chosen in rather broad intervals. In what follows we list these parameters, with short comments.
\begin{itemize}
\item $\lambda$  is any fixed number between $0$ and $1$ (this parameter controls the ratio between the lengths of messages sent by Alice and Bob);
\item $\kappa_1,\kappa_2$ (some absolute constants that control the asymptotic of communication complexity hidden in the $o(\cdot)$-terms in the statements of Theorem~\ref{thm-main} and Proposition~\ref{prop-model-3}); 
\item $k(n)=\log n$ (we will cut strings of Alice and Bob in ``blocks'' of length $k$; we can afford the brute force search over all binary strings of length $k$, since $2^k$ is polynomial in $n$);
\item $m(n) = n/k(n)$ (when we split $n$-bits strings into blocks of length $k$, we get $m$ blocks);
\item $r(n) = O(\log k) = O(\log \log n)$ (this parameter controls the chances to get a collision in hashing; we choose $r(n)$  so that $1\ll r(n) \ll k$);
\item $\delta(n) = k^{-0.49} = (\log n)^{-0.49}$ (the threshold for deviation of the relative frequency from the probability involved in the law of large numbers; notice that we choose $\delta(n)$  such that $ \frac1{\sqrt k} \ll \delta(n)\ll k$);
\item $\sigma = \Theta(\frac{1}{(\log n)^c})$ for some constant $c > 0$ (in our construction $\sigma n$ is the length of the Reed-Solomon checksum; we chose $\sigma$ such that  $\sigma\to 0$);
\item $t$ (this parameter characterize the quality of the random bits used by Alice and Bob; accordingly, this parameter is involved in  the law(s) of large numbers used to bound the probability of the error; we let $t(n)=m^{c}$ for some $c>0$).
\end{itemize}

\subsubsection{The scheme of the protocol}

\noindent
\textbf{Alice's part of the protocol:}

\begin{itemize}
\item[($1_A$)] Select at random a tuple of $\lambda n$ indices $I=\{i_1, i_2, \ldots, i_{\lambda n}\}\subset \{1,\ldots, n\}$. Technically, we may assume that Alice chooses at random a permutation $\pi_I$ on the set $\{1, 2, \ldots, n\}$ and lets $I := \pi_I(\{1,2,\ldots,\lambda n\})$. 
\item[($2_A$)] Send to the receiver the bits $X_I = x_{i_1} \ldots x_{i_{\lambda n}}$, see Fig.~\ref{fig-8}.
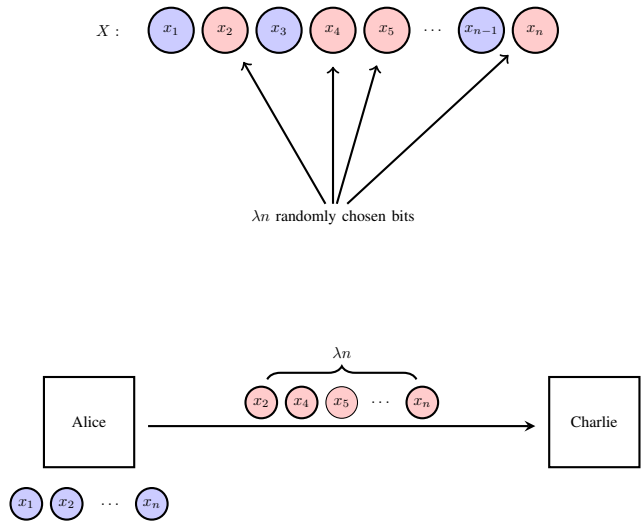
\begin{figure}[t]
\begin{center}
\begin{tikzpicture}[thick,scale=0.6, every node/.style={scale=0.6}]
    \node[] (dummy) {$X:$};
    \node[circle,fill=blue!20,draw,minimum size=1cm,inner sep=0pt] (1) [right = 0.3 cm of dummy]{$x_1$};
    \node[circle,fill=red!20,draw,minimum size=1cm,inner sep=0pt] (2) [right = 0.1cm of 1]  {$x_2$};
    \node[circle,fill=blue!20,draw,minimum size=1cm,inner sep=0pt] (3) [right = 0.1cm of 2]  {$x_3$};
    \node[circle,fill=red!20,draw,minimum size=1cm,inner sep=0pt] (4) [right = 0.1cm of 3]  {$x_4$};
    \node[circle,fill=red!20,draw,minimum size=1cm,inner sep=0pt] (5) [right = 0.1cm of 4]  {$x_5$};
    
    \node[] (6) [right = 0.1cm of 5]  {$\dots$};
    
    \node[circle,fill=blue!20,draw,minimum size=1cm,inner sep=0pt] (7) [right = 0.1cm of 6]  {$x_{n-1}$};
    \node[circle,fill=red!20,draw,minimum size=1cm,inner sep=0pt] (8) [right = 0.1cm of 7]  {$x_n$};
    
    \node[] (9) [below = 2cm of 4] {$\lambda n$ randomly chosen bits};
    
    \draw[->, shorten >= 5pt] (9)--(2);
    \draw[->, shorten >= 5pt] (9)--(4);
    \draw[->, shorten >= 5pt] (9)--(5);
    \draw[->, shorten >= 5pt] (9)--(8);
\end{tikzpicture} 

\rule{0pt}{1.5cm}

\begin{tikzpicture}[>=stealth, thick,scale=0.6, every node/.style={scale=0.6}]
\node[draw, minimum width=2cm, minimum height=2cm] (A) {Alice};
\node[draw, minimum width=2cm, minimum height=2cm, right=5.5cm of A] (C) {Charlie};
\draw[->,thick, shorten >= 5pt, shorten <= 5pt] (A.-5) -- node[thin,above = 0.1, circle,fill=red!20,draw,minimum size=0.7cm,inner sep=0pt] (1) {$x_5$}  (C.-175);
\node[] (2) [right = 0.1cm of 1]  {$\dots$};
\node[circle,fill=red!20,draw,minimum size=0.7cm,inner sep=0pt] (3) [right = 0.1cm of 2]  {$x_n$};
\node[circle,fill=red!20,draw,minimum size=0.7cm,inner sep=0pt] (4) [left = 0.1cm of 1]  {$x_4$};
\node[circle,fill=red!20,draw,minimum size=0.7cm,inner sep=0pt] (5) [left = 0.1cm of 4]  {$x_2$};

\node[] (6) [above = 0.01cm of 5] {}; 
\node[] (7) [above = 0.01cm of 3] {}; 

\node[minimum size = 0] (70) [below = 0.4cm of A] {};
\node[circle,fill=blue!20,draw,minimum size=0.7cm,inner sep=0pt] (8) [left = 0 of 70]  {$x_2$};
\node[circle,fill=blue!20,draw,minimum size=0.7cm,inner sep=0pt] (9) [left = 0.1 of 8]  {$x_1$};
\node[circle,minimum size=0.7cm,inner sep=0pt] (10) [right = 0 of 70]  {$\dots$};
\node[circle,fill=blue!20,draw,minimum size=0.7cm,inner sep=0pt] (11) [right = 0.1 of 10]  {$x_n$};

\draw [decorate,decoration={brace,amplitude=5pt}] (6)  -- (7) 
   node [above=0.2cm,midway] {$\lambda n$};
\end{tikzpicture}
\end{center}
\caption{Steps $1_A$ and $2_A$: Alice selects $\lambda n$ bits in $X$ and sends them to Charlie.}
\label{fig-8}
\end{figure}

\item[($3_A$)] Choose another random permutation $\pi_A : \{1,\ldots,n\} \to \{1,\ldots,n\}$ and permute the bits of $X$, i.e., 
let\footnote{In what follows we consider also the $\pi_A$-permutation of bits in $Y$ and denote it
 $
 Y' = y'_1 \ldots y'_n   := y_{\pi_A(1)} \ldots y_{\pi_A(n)}.
 $
 Thus, the prime in the notation (e.g., $X'$ and $Y'$) implies that we permuted the bits of the original strings by $\pi_A$.}
 $
 X'  = x'_1 \ldots x'_n  := x_{\pi_A(1)} \ldots x_{\pi_A(n)}
 $
(see Fig.~\ref{fig-9}). 
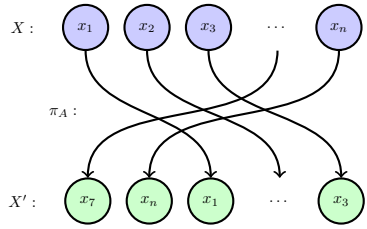
\begin{figure}[t]
\begin{center}
\begin{tikzpicture}[thick,scale=0.6, every node/.style={scale=0.6}]
    \node[] (dummy) {$X:$};
    \node[circle,fill=blue!20,draw,minimum size=1cm,inner sep=0pt] (1) [right = 0.3 cm of dummy]{$x_1$};
    \node[circle,fill=blue!20,draw,minimum size=1cm,inner sep=0pt] (2) [right = 0.2cm of 1]  {$x_2$};
    \node[circle,fill=blue!20,draw,minimum size=1cm,inner sep=0pt] (3) [right = 0.2cm of 2]  {$x_3$};
    
    \node[circle,minimum size=1cm,inner sep=0pt] (4) [right = 0.3cm of 3]  {$\dots$};
    
    \node[circle,fill=blue!20,draw,minimum size=1cm,inner sep=0pt] (5) [right = 0.2cm of 4]  {$x_n$};
    
    \node[below=2 of dummy] (dummy2) {$X':$};
    \node[circle,fill=green!20,draw,minimum size=1cm,inner sep=0pt] (12) [right = 0.3 cm of dummy2]{$x_7$};
    \node[circle,fill=green!20,draw,minimum size=1cm,inner sep=0pt] (22) [right = 0.2cm of 12]  {$x_n$};
    \node[circle,fill=green!20,draw,minimum size=1cm,inner sep=0pt] (32) [right = 0.2cm of 22]  {$x_1$};
    
    \node[circle,minimum size=1cm,inner sep=0pt] (42) [right = 0.3cm of 32]  {$\dots$};
    
    \node[circle,fill=green!20,draw,minimum size=1cm,inner sep=0pt] (52) [right = 0.2cm of 42]  {$x_3$};
    
    \draw (1.-90) edge[out=270,in=90,->] (32.90);
    \draw (2.-90) edge[out=270,in=90,->] (42.90);
    \draw (3.-90) edge[out=270,in=90,->] (52.90);
    \draw (4.-90) edge[out=270,in=90,->] (12.90);
    \draw (5.-90) edge[out=270,in=90,->] (22.90);
    
    \node[] (BT) [below=0.9 of dummy]  {};
    \node[] (P) [right=0.2 of BT]  {$\pi_A:$};
\end{tikzpicture} 
\end{center}
\caption{Step $3_A$: $\pi_A$ permutes the bits of $X=x_1x_2\ldots x_n$ and obtains $X'$}
\label{fig-9}
\end{figure}
 Further, divide $X'$ into blocks of length $k(n)$, i.e., represent $X'$ as a concatenation
   $
   X' = X'_1 \ldots X'_m,
   $
where $X'_j := x'_{(j-1)k+1} x'_{(j-1)k+2} \ldots x'_{jk}$ for each $j$ (see Fig.~\ref{fig-10}). 

\item[($4_A$)] Then Alice computes hash values of these blocks. More technically, we consider a universal family of hash functions 
 $$
  \hash^A_l \ :\ \{0,1\}^k \to \{0,1\}^{h(\alpha) (1-\lambda) k + \kappa_1 \delta k + \kappa_2 \log k+r}.
 $$
With some standard universal hash family, we may assume that these hash functions are indexed  by bit strings $l$ of length $O(k)$, see  Proposition~\ref{proposition-hash}. 
Alice chooses at random $m$  indices $l_1,\ldots, l_m$ of hash functions. 
Then Alice applies each $\hash_{l_j}$ to the corresponding  block $X'_j$   and sends to Charlie  the resulting hash values 
 $
 \hash^A_{l_1}(X_1'), \ldots,  \hash^A_{l_m}(X_m'),
 $
 see Fig.~\ref{fig-10}.
 
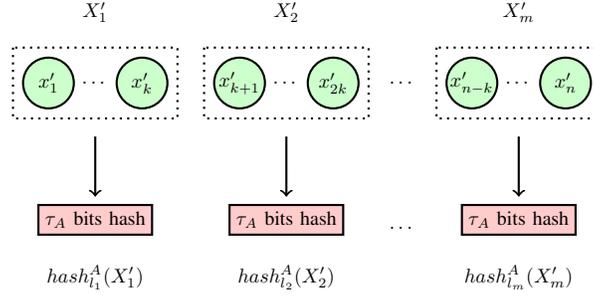
\begin{figure}[t]
\begin{center}
\begin{tikzpicture}[thick,scale=0.75, every node/.style={scale=0.75}]
    \node[circle,draw,fill=green!20,minimum size=0.9cm,inner sep=0pt] (1) {$x'_1$};
    \node[] (20) [right = 0cm of 1]  {$\dots$};
    \node[circle,draw,fill=green!20,minimum size=0.9cm,inner sep=0pt] (3) [right = 0cm of 20]  {$x'_k$};
    
    \node[circle,draw,fill=green!20,minimum size=0.9cm,inner sep=0pt] (4) [right = 0.6cm of 3]  {$x'_{k+1}$};
    \node[] (50) [right = 0cm of 4]  {$\dots$};
    \node[circle,draw,fill=green!20,minimum size=0.9cm,inner sep=0pt] (6) [right = 0cm of 50]  {$x'_{2k}$};
    
    \node[] (7) [right = 0.3cm of 6]  {$\dots$};
    
    \node[circle,draw,fill=green!20,minimum size=0.9cm,inner sep=0pt] (8) [right = 0.3cm of 7]  {$x'_{n-k}$};
    \node[] (90) [right = 0cm of 8]  {$\dots$};
    \node[circle,draw,fill=green!20,minimum size=0.9cm,inner sep=0pt] (10) [right = 0cm of 90]  {$x'_n$};

    \draw[thick, dotted] ($(1.north west)+(-0.3,0.3)$) rectangle ($(3.south east)+(0.3,-0.3)$);
    \draw[thick, dotted] ($(4.north west)+(-0.3,0.3)$) rectangle ($(6.south east)+(0.3,-0.3)$);
    \draw[thick, dotted] ($(8.north west)+(-0.3,0.3)$) rectangle ($(10.south east)+(0.3,-0.3)$);
    
    \node[above = 0.6cm of 20] {$X'_1$};
    \node[above = 0.6cm of 50] {$X'_2$};
    \node[above = 0.6cm of 90] {$X'_m$};

    \node[draw, fill=red!20, below = 1.5cm of 20] (200) {$\tau_A$ bits hash};
    \draw[->, shorten <= 0.6cm, shorten >= 0.1cm] (20)--(200);
    \node[draw, fill=red!20, below = 1.5cm of 50] (500) {$\tau_A$ bits hash};
    \draw[->, shorten <= 0.6cm, shorten >= 0.1cm] (50)--(500);
    \node[below = 1.7cm of 7] {$\dots$};
    \node[draw, fill=red!20, below = 1.5cm of 90] (900) {$\tau_A$ bits hash};
    \draw[->, shorten <= 0.6cm, shorten >= 0.1cm] (90)--(900);
    
    \node[below = 0.3cm of 200] {$hash_{l_1}^A(X'_1)$};
    \node[below = 0.3cm of 500] {$hash_{l_2}^A(X'_2)$};
    \node[below = 0.3cm of 900] {$hash_{l_m}^A(X'_m)$};
\end{tikzpicture} 
\end{center}
\caption{Step $4_A$: hashing the blocks of $X'$; the length of each hash is equal to $\tau_A:=h(\alpha) (1-\lambda) k + \kappa_1 \delta k + \kappa_2 \log k+r$.}
\label{fig-10}
\end{figure}
   
\item[($5_A$)] Compute the Reed-Solomon checksums of the sequence $X'_1, \ldots, X'_m$ that are enough to reconstruct all blocks $X_j'$  if most $\sigma m$ of them are corrupted, and send them to Charlie. These checksums make a string of $O(\sigma mk)$ bits, see Proposition~\ref{proposition-rs}.

\end{itemize}

\noindent
\textbf{Summary:} Alice sends to Charlie three tranches of information,
\begin{itemize}
\item[(i)] $\lambda n$ bits of $X$ selected at random,
\item[(ii)] hashes for each of $m=n/k$ blocks in the permuted string $X'$,
\item[(iii)] the Reed--Solomon checksums for the blocks of $X'$.
\end{itemize}

\medskip

\noindent
\textbf{Bob's part of the protocol:}

\begin{itemize}

\item[($1_B$)] Choose at random permutation $\pi_B : \{1,\ldots,n\} \to \{1,\ldots,n\}$ and use it to permute the bits of $Y$, i.e., 
let\footnote{Similarly, in what follows we apply this permutation to the bits of $X$ and denote
 $
 X'' = x''_1 \ldots x''_n   := x_{\pi_B(1)} \ldots x_{\pi_B(n)}.
 $
 Thus, the double prime in the notation (e.g., $X''$ and $Y''$) implies that we permuted the bits of the original strings by $\pi_B$.}
 $
 Y''  =  y''_1\ldots y''_n  := y_{\pi_B(1)} \ldots y_{\pi_B(n)},
 $
 see Fig.~\ref{fig-11}.
 \begin{figure}[ht]
 \begin{center}
\begin{tikzpicture}[thick,scale=0.6, every node/.style={scale=0.6}]
    \node[] (dummy) {$Y:$};
    \node[circle,fill=blue!20,draw,minimum size=1cm,inner sep=0pt] (1) [right = 0.3 cm of dummy]{$y_1$};
    \node[circle,fill=blue!20,draw,minimum size=1cm,inner sep=0pt] (2) [right = 0.2cm of 1]  {$y_2$};
    \node[circle,fill=blue!20,draw,minimum size=1cm,inner sep=0pt] (3) [right = 0.2cm of 2]  {$y_3$};
    
    \node[circle,minimum size=1cm,inner sep=0pt] (4) [right = 0.3cm of 3]  {$\dots$};
    
    \node[circle,fill=blue!20,draw,minimum size=1cm,inner sep=0pt] (5) [right = 0.2cm of 4]  {$y_n$};
    
    \node[below=2 of dummy] (dummy2) {$Y'':$};
    \node[circle,fill=orange!20,draw,minimum size=1cm,inner sep=0pt] (12) [right = 0.3 cm of dummy2]{$y_4$};
    \node[circle,fill=orange!20,draw,minimum size=1cm,inner sep=0pt] (22) [right = 0.2cm of 12]  {$y_{n-1}$};
    \node[circle,fill=orange!20,draw,minimum size=1cm,inner sep=0pt] (32) [right = 0.2cm of 22]  {$y_1$};
    
    \node[circle,minimum size=1cm,inner sep=0pt] (42) [right = 0.3cm of 32]  {$\dots$};
    
    \node[circle,fill=orange!20,draw,minimum size=1cm,inner sep=0pt] (52) [right = 0.2cm of 42]  {$y_2$};
    
    \draw (1.-90) edge[out=270,in=90,->] (32.90);
    \draw (2.-90) edge[out=270,in=90,->] (42.90);
    \draw (3.-90) edge[out=270,in=90,->] (52.90);
    \draw (4.-90) edge[out=270,in=90,->] (12.90);
    \draw (5.-90) edge[out=270,in=90,->] (22.90);
    
    \node[] (BT) [below=0.9 of dummy]  {};
    \node[] (P) [right=0.2 of BT]  {$\pi_B:$};
\end{tikzpicture} 
\end{center}
\caption{Step $1_B$: $\pi_B$ permutes the bits of $Y=y_1y_2\ldots y_n$ and obtains $Y'$}
\label{fig-11}
 \end{figure}
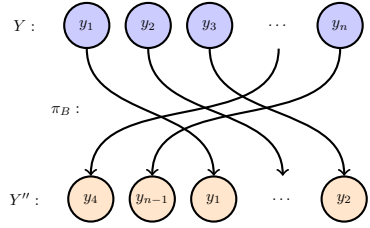

 Further, divide $Y''$ into blocks of length $k$, and represent $Y''$ as a concatenation
   $
   Y'' = Y''_1 \ldots Y''_m,
   $ 
where $Y''_j := y''_{(j-1)k+1} y''_{(j-1)k+2} \ldots y''_{jk}$ for each $j$, see Fig.~\ref{fig-12}.

\item[($2_B$)] Then choose at random $m$ hash functions $\hash^B_{l_j}$ from a universal family of hash functions 
 $$
  \hash^B_l  \,:\, \{0,1\}^k \to \{0,1\}^{(1-\lambda)k + h(\alpha) \lambda k + \kappa_1 \delta k + \kappa_2\log k+r}
 $$
(we assume that $l_j$ are $(T/k)$-independent) and send to Charlie random hash values 
 $
 \hash^B_{l_1}(Y_1''), \ldots,  \hash^B_{l_m}(Y_m''),
 $
 see Fig.~\ref{fig-12}.
 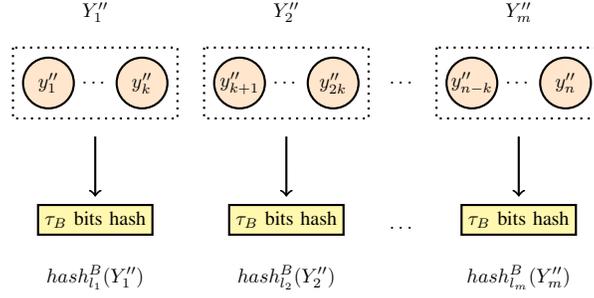
\begin{figure}[t]
 \begin{center}
\begin{tikzpicture}[thick,scale=0.75, every node/.style={scale=0.75}]
    \node[circle,draw,fill=orange!20,minimum size=0.9cm,inner sep=0pt] (1) {$y''_1$};
    \node[] (20) [right = 0cm of 1]  {$\dots$};
    \node[circle,draw,fill=orange!20,minimum size=0.9cm,inner sep=0pt] (3) [right = 0cm of 20]  {$y''_k$};
    
    \node[circle,draw,fill=orange!20,minimum size=0.9cm,inner sep=0pt] (4) [right = 0.6cm of 3]  {$y''_{k+1}$};
    \node[] (50) [right = 0cm of 4]  {$\dots$};
    \node[circle,draw,fill=orange!20,minimum size=0.9cm,inner sep=0pt] (6) [right = 0cm of 50]  {$y''_{2k}$};
    
    \node[] (7) [right = 0.3cm of 6]  {$\dots$};
    
    \node[circle,draw,fill=orange!20,minimum size=0.9cm,inner sep=0pt] (8) [right = 0.3cm of 7]  {$y''_{n-k}$};
    \node[] (90) [right = 0cm of 8]  {$\dots$};
    \node[circle,draw,fill=orange!20,minimum size=0.9cm,inner sep=0pt] (10) [right = 0cm of 90]  {$y''_n$};

    \draw[thick, dotted] ($(1.north west)+(-0.3,0.3)$) rectangle ($(3.south east)+(0.3,-0.3)$);
    \draw[thick, dotted] ($(4.north west)+(-0.3,0.3)$) rectangle ($(6.south east)+(0.3,-0.3)$);
    \draw[thick, dotted] ($(8.north west)+(-0.3,0.3)$) rectangle ($(10.south east)+(0.3,-0.3)$);
    
    \node[above = 0.6cm of 20] {$Y''_1$};
    \node[above = 0.6cm of 50] {$Y''_2$};
    \node[above = 0.6cm of 90] {$Y''_m$};

    \node[draw, fill=yellow!40, below = 1.5cm of 20] (200) {$\tau_B$ bits hash};
    \draw[->, shorten <= 0.6cm, shorten >= 0.1cm] (20)--(200);
    \node[draw, fill=yellow!40, below = 1.5cm of 50] (500) {$\tau_B$ bits hash};
    \draw[->, shorten <= 0.6cm, shorten >= 0.1cm] (50)--(500);
    \node[below = 1.7cm of 7] {$\dots$};
    \node[draw, fill=yellow!40, below = 1.5cm of 90] (900) {$\tau_B$ bits hash};
    \draw[->, shorten <= 0.6cm, shorten >= 0.1cm] (90)--(900);
    
    \node[below = 0.3cm of 200] {$hash_{l_1}^B(Y''_1)$};
    \node[below = 0.3cm of 500] {$hash_{l_2}^B(Y''_2)$};
    \node[below = 0.3cm of 900] {$hash_{l_m}^B(Y''_m)$};
\end{tikzpicture} 
\end{center}
\caption{Step $2_B$: hashing the blocks of $X'$; the length of each hash is equal to $\tau_B:=(1-\lambda)k + h(\alpha) \lambda k + \kappa_1 \delta \cdot k + \kappa_2\log k+r$.}
\label{fig-12}
 \end{figure}
 
Similarly to ($4_A$), we may assume that these hash functions are indexed  by bit strings $l$ of length $O(k)$, see  Proposition~\ref{proposition-hash}.

\item[($3_B$)] Compute the Reed-Solomon checksums of the sequence $Y''_1, \ldots, Y''_m$, that are enough to reconstruct all blocks $Y_j''$, if at most $\sigma m$ of them are corrupted, and send them to Charlie. These checksums should be a string of length $O(\sigma m k)$ bits,  see Proposition~\ref{proposition-rs}.
\end{itemize}

\noindent
\textbf{Summary:} Bob sends to Charlie two tranches of information,
\begin{itemize}
\item[(i)] hashes for each of $m=n/k$ blocks in the permuted string $Y''$,
\item[(ii)] the Reed--Solomon checksums for the blocks of $Y''$.
\end{itemize}

\medskip

\noindent
\textbf{Charlie's part of the protocol:}

\begin{itemize}
\item[($1_C$)] Apply Bob's permutation  $\pi_B$ to the positions of bits selected by Alice, and denote the result by $I''$,  i.e.,
 $$
 I'' = \{\pi_B(i_1), \ldots, \pi_B(i_{\lambda n}) \}.
 $$
Then split indices of $I''$ into $m$ disjoint parts corresponding to the different intervals $Int_j = \{(j-1)k+1, (j-1)k+2,  \ldots, jk \}$, and
  $I''_j := I'' \cap Int_j$ (the red nodes in Fig.~\ref{fig-13}).
Further, for each $j=1,\ldots,m$   denote by $X_{I''_j}$ the bits sent by Alice, that appear in the interval $Int_j$  after permutation $\pi_B$.
 (We will show later that the typical size of  $X_{I''_j}$ is close to $\lambda k$.) 
  \begin{figure}[ht]
\begin{center}
\begin{tikzpicture}[thick,scale=0.75, every node/.style={scale=0.75}]

    \node[] (X) {$X''_j:$};
    \node[] (Y) [below = 2cm of X] {$Y''_j:$};

    \node[] (1) [right = 0.5cm of X] {$\dots$};
    
    \node[circle,draw,fill=red!20,minimum size=0.9cm,inner sep=0pt] (4) [right = 0.6cm of 1]  {$x_{42}$};
    \node[circle,draw,fill=orange!20,minimum size=0.9cm,inner sep=0pt] (5) [right = 0.1cm of 4]  {$?$};
    \node[circle,draw,fill=orange!20,minimum size=0.9cm,inner sep=0pt] (51) [right = 0.1cm of 5]  {$?$};
    \node[circle,draw,fill=red!20,minimum size=0.9cm,inner sep=0pt] (52) [right = 0.1cm of 51]  {$x_{357}$};
    \node[circle,draw,fill=orange!20,minimum size=0.9cm,inner sep=0pt] (53) [right = 0.1cm of 52]  {$?$};
    
    \node[] (50) [right = 0cm of 53]  {$\dots$};
    \node[circle,draw,fill=red!20,minimum size=0.9cm,inner sep=0pt] (6) [right = 0cm of 50]  {$x_{73}$};
    
    \node[] (7) [right = 0.6cm of 6]  {$\dots$};

    \draw[thick, dotted] ($(4.north west)+(-0.3,0.3)$) rectangle ($(6.south east)+(0.3,-0.3)$);
    
    \node[] (666) [above = 1.2cm of 52] {$\approx\lambda k$ bits are received from Alice};
    
    \draw[->, shorten >= 2pt] (666)--(4);
    \draw[->, shorten >= 2pt] (666)--(52);
    \draw[->, shorten >= 2pt] (666)--(6);

    \node[] (1Y) [right = 0.5cm of Y] {$\dots$};
    
    \node[circle,draw,fill=orange!20,minimum size=0.9cm,inner sep=0pt] (4Y) [right = 0.6cm of 1Y]  {$?$};
    \node[circle,draw,fill=orange!20,minimum size=0.9cm,inner sep=0pt] (5Y) [right = 0.1cm of 4Y]  {$?$};
    \node[circle,draw,fill=orange!20,minimum size=0.9cm,inner sep=0pt] (51Y) [right = 0.1cm of 5Y]  {$?$};
    \node[circle,draw,fill=orange!20,minimum size=0.9cm,inner sep=0pt] (52Y) [right = 0.1cm of 51Y]  {$?$};
    \node[circle,draw,fill=orange!20,minimum size=0.9cm,inner sep=0pt] (53Y) [right = 0.1cm of 52Y]  {$?$};
    
    \node[] (50Y) [right = 0cm of 53Y]  {$\dots$};
    \node[circle,draw,fill=orange!20,minimum size=0.9cm,inner sep=0pt] (6Y) [right = 0cm of 50Y]  {$?$};
    
    \node[] (7Y) [right = 0.6cm of 6Y]  {$\dots$};

    \draw[thick, dotted] ($(4Y.north west)+(-0.3,0.3)$) rectangle ($(6Y.south east)+(0.3,-0.3)$);
    

\end{tikzpicture} 
\end{center}
\caption{Step $6_C$: for each block $X''_j$ Charlie typically gets from Alice $\approx \lambda k$ bits}
\label{fig-13}
\end{figure}
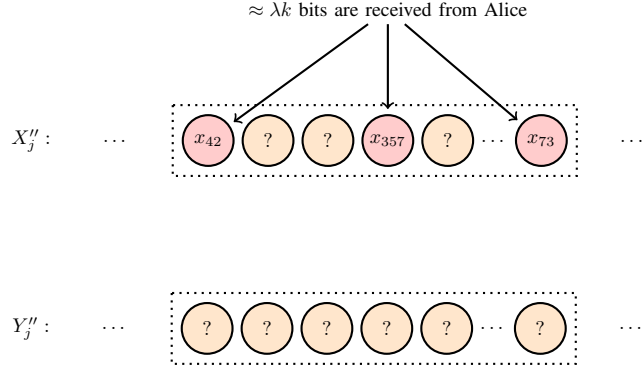

\item[($2_C$)]  For each $j=1,\ldots, m$  try to reconstruct $Y''_j$.  To this end, find all  bit strings
$Z = z_1\ldots z_k$ that satisfy a pair of conditions ($\mathrm{Cond}_1$) and ($\mathrm{Cond}_2$) that we formulate below. 

We abuse notation and denote by $Z_{I''_j}$ the subsequence of bits from $Z$ that appear at the  positions determined by $I''_j$. That is, if 
$$I''_j = \{(j-1)k + s_1,   \cdots,  (j-1)k + s_l \},$$ where
 $$
  (j-1)k + s_1 <  (j-1)k + s_2 < \cdots < (j-1)k + s_l,
 $$
then $Z_{I''_j} = z_{s_1} z_{s_2} \ldots z_{s_l}$.  With  this notation we can specify the required property of $Z$:
 \begin{itemize}
 \item[($\mathrm{Cond}_1$)]   $\dist (X_{I''_j}, Z_{I''_j}) \le (\alpha + \delta) |I''_j|$,  see Fig.~\ref{fig-14},
  \item[($\mathrm{Cond}_2$)]  $ \hash^B_{l_j}(Z)$ must coincide with the hash value $ \hash^B_{l_j}(Y''_j)$ received from Bob.
 \end{itemize}
 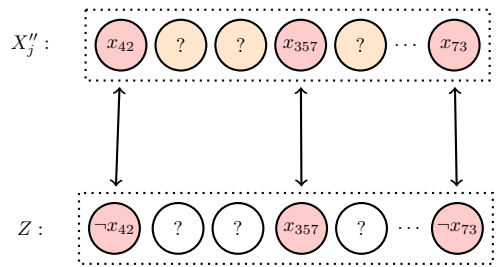
\begin{figure}[t]
\begin{center}
\begin{tikzpicture}[thick,scale=0.75, every node/.style={scale=0.75}]

    \node[] (X) {$X_j'':$};
    \node[] (Y) [below = 2cm of X] {$Z:$\ };
    
    \node[circle,draw,fill=red!20,minimum size=0.9cm,inner sep=0pt] (4) [right = 0.5cm of X]  {$x_{42}$};
    \node[circle,draw,fill=orange!20,minimum size=0.9cm,inner sep=0pt] (5) [right = 0.1cm of 4]  {$?$};
    \node[circle,draw,fill=orange!20,minimum size=0.9cm,inner sep=0pt] (51) [right = 0.1cm of 5]  {$?$};
    \node[circle,draw,fill=red!20,minimum size=0.9cm,inner sep=0pt] (52) [right = 0.1cm of 51]  {$x_{357}$};
    \node[circle,draw,fill=orange!20,minimum size=0.9cm,inner sep=0pt] (53) [right = 0.1cm of 52]  {$?$};
    
    \node[] (50) [right = 0cm of 53]  {$\dots$};
    \node[circle,draw,fill=red!20,minimum size=0.9cm,inner sep=0pt] (6) [right = 0cm of 50]  {$x_{73}$};

    \draw[thick, dotted] ($(4.north west)+(-0.3,0.3)$) rectangle ($(6.south east)+(0.3,-0.3)$);
    
    \node[circle,draw,fill=red!20,minimum size=0.9cm,inner sep=0pt] (4Y) [right = 0.5cm of Y]  {$\neg x_{42}$};
    \node[circle,draw,minimum size=0.9cm,inner sep=0pt] (5Y) [right = 0.15cm of 4Y]  {$?$};
    \node[circle,draw,minimum size=0.9cm,inner sep=0pt] (51Y) [right = 0.1cm of 5Y]  {$?$};
    \node[circle,draw,fill=red!20,minimum size=0.9cm,inner sep=0pt] (52Y) [right = 0.15cm of 51Y]  {$x_{357}$};
    \node[circle,draw,minimum size=0.9cm,inner sep=0pt] (53Y) [right = 0.1cm of 52Y]  {$?$};
    
    \node[] (50Y) [right = 0.03cm of 53Y]  {$\dots$};
    
    \node[circle,draw,fill=red!20,minimum size=0.9cm,inner sep=0pt] (6Y) [right = 0cm of 50Y]  {$\neg x_{73}$};

    \draw[thick, dotted] ($(4Y.north west)+(-0.3,0.3)$) rectangle ($(6Y.south east)+(0.3,-0.3)$);

   

    \draw[<->, shorten >= 6pt, shorten <= 6pt] (4)--(4Y);
    \draw[<->, shorten >= 6pt, shorten <= 6pt] (52)--(52Y);
    \draw[<->, shorten >= 6pt, shorten <= 6pt] (6)--(6Y);
\end{tikzpicture} 
\end{center}
\caption{Alice sends to Charlie  $\approx \lambda k$ bits of the block $X''_j$ (red nodes);  to construct a $Z$, Charlie inverts $\approx \alpha\cdot \lambda k$ bits of $x_i$ in the ``red'' positions, and then choose the other bits arbitrarily.}
\label{fig-14}
\end{figure}

If there is a unique   $Z$ that satisfies these two conditions, then take it as a candidate for $Y''_j$;
otherwise (if there is no such $Z$ or if there exist more than one $Z$ that satisfy these conditions) 
we say that the procedure of reconstruction of $Y''_j$ fails.

\smallskip

\emph{Remark 1:} The requirement from~($\mathrm{Cond}_1$) makes sense since   in a typical case the indices from $I''$ are somehow uniformly distributed.
  
 \smallskip
  
\emph{Remark 2:} There can be  two kinds of troubles at this stage. First, for some blocks $Y''_j$ reconstruction fails (this happens when Charlie gets no or more than one  $Z$ that satisfy~($\mathrm{Cond}_1$) and ~($\mathrm{Cond}_2$)). Second, for some blocks $Y''_j$ the reconstruction procedure seemingly completes, but the obtained result is incorrect (Charlie gets a $Z$ which is not the real value of $Y''_j$). In what follows we prove that both events are rare. In a typical case, at this stage  most (but not all!) blocks $Y''_j$ are correctly reconstructed.

 \item[($3_C$)] Use Reed-Solomon checksums received from Bob to correct the blocks $Y''_j$ that we failed to reconstruct or reconstructed incorrectly  at step~($2_C$).
 
\emph{Remark 3:}   Below we prove that in a typical case, after this procedure we get correct values of all blocks $Y''_j$, so concatenation of these blocks gives $Y''$.

 \item[($4_C$)] Apply permutation $\pi^{-1}_B$ to the bits of $Y''$ and obtain $Y$.
 
 \item[($5_C$)]  Permute bits of $Y$ and $X_I$ using permutation $\pi_A$. 
 
  \item[($6_C$)]  For each $j=1,\ldots, m$  try to reconstruct  $X'_j$. 
To this end,  find all   bit strings $W =w_1\ldots w_k$ such  that 
 \begin{itemize}
  \item[($\mathrm{Cond}_3$)]  at each position  from 
   $I' \cap Int_j$
  the bit from $X'$ (in the $j$-th block) sent by Alice coincides with the corresponding bit in $W$,
  \item[($\mathrm{Cond}_4$)]   $\dist (Y'_{Int_j \setminus I'_j}, W_{Int_j \setminus I'_j}) \le (\alpha + \delta) |Int_j \setminus I'_j|$
 \item[($\mathrm{Cond}_5$)] $ \hash^A_{l_j}(W)$ coincides with the hash value $ \hash^A_{l_j}(X'_j)$ received from Alice.
 \end{itemize}
If there is a unique  $W$ that satisfies these conditions, then take this string as a candidate for $X'_j$;
otherwise (if there is no such $W$ or if there exist more than one $W$ satisfying these conditions) we say that reconstruction of $X'_j$ fails. 

\emph{Remark:} We will show that in a typical case, most (but not all) blocks $X'_j$ will be correctly reconstructed.

\item[($7_C$)] Use Reed-Solomon checksums received from Alice to correct the blocks $X'_j$ that were incorrectly decoded at step~($6_C$).

\emph{Remark:}   We will show in a typical case, after this procedure we get correct values of all blocks $X'_j$, so concatenation of these blocks gives $X'$.

\item[($8_C$)] Apply permutation $\pi^{-1}_A$ to the positions of bits of $X'$ and obtain $X$.

\end{itemize}

The main technical result of this section (correctness of the protocol) follows from the next lemma.
\begin{lemma}\label{lemma-error-model-2}
In Communication Model~1, the protocol described above fails with probability at most $O(2^{-m^d})$ for some $d>0$.
\end{lemma}
(The proof is deferred to Section~\ref{appendix}.)

\subsubsection{Communication complexity of the protocol.}

Alice sends $\lambda n$ bits at step~($2_A$), $h(\alpha) (1- \lambda) k + O(\delta ) k + O(\log k) +r $ for each block $j=1,\ldots, m$ at step~($3_A$), and $\sigma m k$ bits of the Reed-Solomon checksums at step~($4_A$). So the total length of Alice's message is 
 $$
    \lambda n + \left( h(\alpha) (1- \lambda) k + O(\delta ) k + O(\log k) +r  \right)\cdot m  + \sigma n.
 $$
For the values of parameters that we have chosen above (see Section~\ref{section-parameters}), this sum can be estimated as 
 $
     \lambda n + h(\alpha) (1 -\lambda) n + o(n).
 $
  \medskip
  Bob sends $ (1-\lambda) k +  h(\alpha) \lambda k  + O(\delta ) k + O(\log k) + r$ bits for each block $j=1,\ldots, m$ at step~($1_B$)
 and $\sigma m k$ bits of the Reed-Solomon checksums at step~($2_B$). This sums up to
  $$
  \left(
  (1-\lambda) k +  h(\alpha) \lambda k  + O(\delta ) k + O(\log k) + r
  \right)\cdot m + \sigma n
  $$
bits.  For the chosen values of parameters this sum is equal to 
$
  (1-\lambda) n +  h(\alpha) \lambda n + o(n). 
$
When we vary parameter  $\lambda$ between $0$ and $1$, we variate accordingly  the lengths of both messages from  $h(\alpha)  n + o(n)$ to $(1+h(\alpha))n+o(n)$,  whereas the sum of Alice's and Bob's messages always remains equal to
$
 (1+ h(\alpha)) n  + o(n).
$
Thus, varying $\lambda$ from $0$ to $1$,  we move in the graph in Fig.~\ref{fig-7}  from $P_B$ to $P_A$.

It remains to notice that algorithms of all participants require only $\poly(n)$-time computations. Indeed, all manipulations with Reed-Solomon checksums (encoding and error-correction) can be done in time $\poly(n)$, with standard encoding and decoding algorithms.
The brute force search used in the decoding procedure requires only the search over sets of size $2^k= \poly(n)$). 
Thus, Proposition~\ref{prop-model-2} is proven.

\subsection{An effective protocol for Model~2}
\label{section-protocol-model-2}

In this section we prove that the pairs of rates from Fig.~\ref{fig-7} are achievable for Communication Model~2.
Now the random sources of Alice and Bob are not perfect: the random permutations are only $t$-wise almost independent and the chosen hash functions are $t$-independent (for a suitable $t$). 
\begin{proposition}\label{prop-model-3}
The version of Theorem~\ref{thm-main} holds for  Communication Model~2 (with parameter $T=\Theta(n^{c}\log n)$).
\end{proposition}
To prove Proposition~\ref{prop-model-3} we do not need a new communication protocol --- in fact, the protocol that we constructed for Model~1 in the previous section works for Model~2 as well. The only difference between  Proposition~\ref{prop-model-2} and Proposition~\ref{prop-model-3} is a more general statement about the estimation of the error probability:

\begin{lemma}\label{lemma-error-model-3}
For Communication Model~2 with parameter $T=\Theta(n^{c}\log n)$
the communication protocol described in section~\ref{section-model-2} fails with probability at most $O(2^{-m^d})$ for some $d>0$.
\end{lemma}
(The proof is deferred to Section~\ref{appendix}.)

Since the protocol remains the same, the bounds for the communication and computational complexity, proven in Proposition~\ref{prop-model-2}, remain valid in the new setting. With Lemma~\ref{lemma-error-model-3} we get the proof of Proposition~\ref{prop-model-3}.

\subsection{The model with private sources of perfect randomness}

Proposition~\ref{prop-model-3} claims that the protocol from Section~\ref{section-model-2} works well for the artificial Communication Model~2 (with non-perfect and partially private randomness).  Now we want to modify this protocol and adapt it to Communication Model~3.  

Technically, we have to get rid of (partially) shared randomness. That is, in Model~3 we cannot assume that Charlie access Alice's and Bob's random bits for free.  Moreover, Alice and Bob cannot just send their random bits to Charlie (this would dramatically increase the communication complexity).  However,  we can use the following well-known trick: we require now that Alice and Bob use pseudo-random bits  instead of truly uniformly random bits.  Alice and Bob take  short seeds for pseudo-random generators at random (with the truly uniform distribution) expand them to longer sequences of pseudo-random bits, and  feed these \emph{pseudo-random} bits in the protocol described in the previous sections. Alice and Bob transmit  the random seeds of their generators to Charlie (the seeds are rather short, so they do not increase communication complexity substantially);  so Charlie  (using the same pseudo-random generators) expands the seeds to the same long pseudo-random sequences and plug them into his side of the protocol.

More formally, we modify the communication protocol described in Section~\ref{section-protocol-model-1}. Now Alice and Bob begin the protocol with the following steps:
\begin{itemize}
\item[($0_A$)] Alice chooses at random the seeds for pseudo-random generators and send them to Charlie. 
\end{itemize}
\begin{itemize}
\item[($0_B$)] Bob chooses at random the seeds for pseudo-random generators and also send them to Charlie. 
\end{itemize}
When these  preparations are done, the protocol proceeds exactly as in Section~\ref{section-protocol-model-1} (steps ($1_A$)--($5_A$) for Alice, ($1_B$)--($3_B$) for Bob, and ($1_C$)--($8_C$) for Charlie). The only difference that all \emph{random} objects (random hash functions and random permutations) are now \emph{pseudo-random}, produces by pseudo-random generators from the chosen  random seeds.

It remains to choose some specific pseudo-random generators that suits our plan. We need two different pseudo-random generators --- one to generate indices of hash functions and another to generate permutations. Constructing a suitable sequence  of pseudo-random hash-functions is simple. Both Alice and Bob needs $m$ random indices $l_i$ of hash functions, and the size of each family of hash functions is  $2^{O(k)} = 2^{O(\log n)}$.   
We need  the property of $t$-independency of $l_i$ for $t=m^{c}$ (for a small enough $c$). To generate these bits we can take a random polynomial of degree  at most $t-1$ over $\mathbb{F}_{2^{O(\log n)}}$. The seed of this ``generator''  is just the tuple of all coefficients of the chosen polynomial, which requires $O(t\log n) = o(n)$ bits. The outcome of the generator (the resulting sequence of pseudo-random bits) is the sequence of values of the chosen polynomial at (some fixed in advance) $m$ different points of the field. The property of $t$-independence follows immediately from the construction: for a randomly chosen polynomial of degree at most $t-1$ the values at any  $t$ points of the field are independent.

The construction of a pseudo-random permutation is  more involved. We use the construction of a pseudo-random permutation from \cite{knr05}. We need the property of $t$-wise almost independence; by Proposition~\ref{proposition-knr}
such a permutation can be effectively produced by a pseudo-random generator with a seed of length $O(t\log n)$.  Alice and Bob chose seeds for all required pseudo-random permutations at random, with the uniform distribution. 

The seeds of the  generators involved in our protocol are much shorter than $n$, so Alice and Bob can send them to Charlie without essentially increasing communication complexity. 

The probability of the error remains the same is in Section~\ref{section-protocol-model-2}, since we plugged in the protocol the pseudo-random bits which are $T$-wise independent. Hence, we can use the bound from Proposition~\ref{prop-model-3}. This concludes the proof of Theorem~\ref{thm-main}.

\subsection{Proofs of the probabilistic lemmas}
\label{appendix}

In this section we prove the technical probabilistic propositions used  to estimate the probability of the failure in our communication protocols.
\begin{IEEEproof}[Proof of Lemma~\ref{lemma1} (a)]
First we prove the statement for a uniformly independent permutations. 
Let $I=\{i_1,\ldots, i_{\rho n}\}$. We denote
$$
\xi_s = \left\{
\begin{array}{rl}
1, & \mbox{if } \pi(i_s) \in \Delta,\\
0, & \mbox{otherwise}.
\end{array}
\right.
$$
We use the fact that the variables $\xi_s$ are ``almost independent''.
Since the permutation $\pi$ is chosen uniformly, we have $\prob[\xi_s = 1] = |\Delta | / n = k/n$ for each $s$. Hence, $\E (\sum \xi_s) = \rho k$. Let us estimate the variance of this random sum.

For $s_1\not= s_2$ we have
 $$
 \prob[\xi_{s_1} = \xi_{s_2}=1] = \frac{k}{n} \cdot \frac{k-1}{n-1} = \left(\frac{k}{n}\right)^2 + O(k/n^2).
 $$
So, the correlation between every two $\xi_s$ is very weak.  We get
\begin{align*}
 \var\left( \sum \xi_s    \right)  &=  \E \left( \sum \xi_s    \right)^2 - \left(  \E \sum \xi_s     \right)^2 \\
 &= \sum_s \E \xi_s   +\sum_{s_1\not=s_2} \E \xi_{s_1} \xi_{s_2} -  \left(  \E \sum_s \xi_s     \right)^2 \\&=O(k).
\end{align*}

Now we apply Chebyshev's inequality 
\begin{equation}
 \prob_{\pi}\big[\  \big| \sum \xi_s  - \rho k \big| >  \delta k\ \big] < \frac{\var (\sum \xi_s)}{(\delta k)^2}  
 = O\left(\frac1{\delta^2 k}\right),
 \label{eq-lemma4-a}
\end{equation}
and we are done.

For a $k$-wise almost independent permutation we should add to the right-hand side of~(\ref{eq-lemma4-a}) the term $O(2^{-k})$, which does not affect the asymptotic of the final result.
\end{IEEEproof}

Before we prove Lemma~\ref{lemma1}~(b), let us formulate a corollary of Lemma~\ref{lemma1}~(a).

\begin{corollary}\label{lemma-a-prim}
Let  $\Delta_1, \ldots,  \Delta_t$ be some disjoint subsets in $ \{1, \ldots,  n \}$ such that $|\Delta_j|=k$ for each $j$.  Then
 for  a uniformly random or $(kt)$-wise almost independent  permutation  
$\pi : \{1,\ldots, n\} \to \{1,\ldots, n\}$,
 $$
 \prob_\pi\big[\   \big| \pi( I ) \cap \Delta_j \big| > (\rho+\delta) k\   \mbox{ for all }j  \big]   \le \mu^t
 $$
  and 
 $$
 \prob_\pi\big[\   \big| \pi( I ) \cap \Delta_j \big| < (\rho-\delta) k\   \mbox{ for all }j  \big]   \le \mu^t.
 $$
\end{corollary}
\begin{IEEEproof}[Proof of Corollary] (sketch) For a uniform permutation it is enough to notice that the events ``\emph{there are too few $\pi$-images of $I$ in $\Delta_j$}'' are negatively correlated with each other. That is, if we denote  
  $$
  E_j := \left\{ \pi\ \Big|\ \big| \pi( I ) \cap \Delta_{j_1} \big| < (\rho -\delta(n))k \right\},
  $$
then
\begin{align*}
 \prob_{\pi}\big[\  E_{j_1}\  \big] > 
 \prob_{\pi}\big[\   E_{j_1}\   \Big|  \  E_{j_2} \   \ \big] >
\prob_{\pi}\big[\   E_{j_1}\   \Big|  \  E_{j_2} \ \mbox{ and } E_{j_3}  \ \big] > \ldots
\end{align*}
It remains to use the bound from~(a) for the unconditional probabilities.

Similarly to the proof of Lemma~\ref{lemma1}~(a), in the case of almost independent permutations the difference of probabilities is negligible.
\end{IEEEproof}

\begin{IEEEproof}[Proof of Lemma~\ref{lemma1} (b)]
Follows immediately from the Corollary~\ref{lemma-a-prim} and Proposition~\ref{proposition-lln}.
\end{IEEEproof}

\begin{IEEEproof}[Proof of Lemma~\ref{lemma-error-model-2} and Lemma~\ref{lemma-error-model-3}]
We prove directly the statement of Lemma~\ref{lemma-error-model-3} (which implies of course Lemma~\ref{lemma-error-model-2}). Let us estimate probabilities of errors at each step of Charlie's part of the protocol.

Step ($1_C$): No  errors.

Step ($2_C$): We should estimate probabilities of errors in reconstructing each block $Y_j''$.

\emph{1st type error:} the number of Alice's bits $x_{i_1},\ldots, x_{i_{\lambda n}}$ that appear in the block $Y''_j$  is less than $(\lambda -\delta)k$. Technically, this event itself is not an error of decoding; but it  is undesirable: we cannot  guarantee the success of reconstruction of $Y''_j$ if we get in this slot too few bits from Alice. Denote the probability of this event (for a block $j=1,\ldots, m$) by $p^B_1$. By the law of large numbers, $p^B_1\to 0$ if $\delta\gg 1/\sqrt{k}$. This fact follows from Lemma~\ref{lemma1}(a).
 
 \medskip
 
 \emph{2nd type error:} $1$st type error does not occur but 
$$\dist (X_{I''_j}, Y_{I''_j}) > (\alpha + \delta) |I''_j|.$$ 
Denote the probability of this event (for a block $j=1,\ldots, m$) by $p^B_2$. Again, by the law of large numbers, $p^B_2\to 0$ if $\delta \gg 1/\sqrt{k}$.  Technically, we apply Lemma~\ref{lemma1}(a) with $\rho = \alpha$ and $\delta=1/k^{0.49}$. 

\emph{3rd type error:} 1st and 2nd type errors do not occur but there exist at least two different strings $Z$ satisfying~(1) and~(2).
We choose the length of hash values for $\hash^B_l$ so that this event happens with probability less than $p^B_3 = 2^{-r}$.
Let us explain this in more detail.

All the positions of $Int_j$ are split into two classes:  the set $I_j''$ and its complement $Int_j \setminus I_j''$. For each position in $I''_j$ Charlie knows the corresponding bit from $X''$ sent by Alice. To  get $Z$, we should 
\begin{itemize}
\item[(i)] invert at most $(\alpha + \delta) |I''_j|$ of Alice's bits (here we use the fact that the 2nd type error does not occur), and

\item[(ii)] choose some bits for the positions  in $Int_j\setminus I_j''$ (we have no specific restrictions for these bits).
\end{itemize}
The number of all strings that satisfy (i) and (ii) is equal to 
\begin{align*}
S_B := \sum\limits_{s=0 }^{(\alpha + \delta) | I''_j | }
 {{| I''_j |} \choose {s}} \cdot 2^{ |Int_j \setminus I''_j|} 
 = 2^{h(\alpha)\lambda  k + (1-\lambda)k + O(\delta) k + O(\log k)}. 
\end{align*}
(In the last equality we use the assumption that the 1st type error  does not occur, so $|Int_j \setminus I''_j| \le (1-\alpha+\delta)k$.)
We set  the length of the hash function $\hash^B_l$ to 
$$ L_B = \log S_B + r  = (1-\lambda)k + h(\alpha) \lambda k + \kappa_1 \delta \cdot k + \kappa_2\log k+r$$
(here we  choose suitable values of parameters $\kappa_1$ and $\kappa_2$). Hence, from Proposition~\ref{proposition-hash} it follows that the probability of the 3rd type error is at most $1/2^r$.

 \medskip
 
We say that a block $Y_j$ is \emph{reconstructible}, if the errors of type $1$, $2$, and $3$ do not occur for this $j$. For each block $Y_j''$, probability to be non-reconstructible is at most $p^B_1 +  p^B_2 + p^B_3$. This sum can be bounded by some threshold $\mu_B=\mu_B(n)$, where $\mu_B(n)\to 0$.  For the chosen parameters $\delta(n)$ and $r(n)$ we have $\mu_B(n) = 1/(\log n)^c$ for some $c>0$.
  
Since for each $j=1,\ldots,m$ the probability that $Y''_j$ is non-reconstructible is less than $\mu$, we conclude that the expected number of non-reconstructible blocks is less than $\mu m$. This is already good news, but we need a stronger statement --- we want  to conclude that with high probability the number of non-reconstructible  blocks is not far above  the expected value. 

Since random permutations in the construction are $(m^c \cdot k)$-wise almost independent and the indices of hash functions are $m^c$-independent, we can apply Proposition~\ref{proposition-lln} and Lemma~\ref{lemma1}(b) and bound the probability   
that the fraction of non-reconstructible blocks is greater than $3\mu_B$. We conclude that this probability is  $ O(2^{-m^{c}})$
for some $c>0$.

We conclude that on  stage~($2_C$) with probability $1-O(2^{-m^{c}})$ Charlie decodes  all blocks of $Y''_j$ except for at most $3\mu_B(n) \cdot m$ of them.

($3_C$) Here Charlie reconstructs the string $Y''$, if the number of  non-re\-con\-struc\-tible blocks $Y''_j$ (at the previous step) is less than 
$3\mu_B(n) \cdot m$. Indeed,  $3\mu_B(n) \cdot m$ is just the number of errors that can be corrected by the Reed-Solomon checksums. Hence, the probability of failure at this step is less than  $O(2^{-m^{c}})$. Here we choose the value of $\sigma$: we let $\sigma=3\mu$.

Steps ($4_C$) and~($5_C$): No  errors.

Step ($6_C$) is similar to step ($2_C$). We need to estimate the probabilities of errors in the reconstruction procedures for each block $X_j'$.

\noindent
\emph{1st type error:} the number of Alice's bits $x_{i_1},\ldots, x_{i_{\lambda n}}$  is less than $(\lambda -\delta)k$. (We cannot guarantee  a correct reconstruction of a block $X'_j$ if there are too few bits from Alice in this slot). We denote the probability of this event by $p^A_1$. From Lemma~\ref{lemma1}(a) it follows that $p^A_1\to 0$ since $\delta = 1/k^{0.49}\gg 1/\sqrt{k}$.

\noindent
\emph{2nd type error:} 1st type error does not occur but 
$$\dist (X'_{Int_j \setminus I'_j }, Y'_{Int_j \setminus I'_j }) > (\alpha + \delta)  | Int_j \setminus I'_j |.$$ 
Denote the probability of this event by $p^A_2$. Again, from Lemma~\ref{lemma1}(a) it follows that $p^A_2\to 0$ since $\delta\gg 1/\sqrt{k}$. 

 \medskip

\noindent
\emph{3rd type error:} 1st and 2nd type errors do not occur but there exist at least two different strings $W$ satisfying~($\mathrm{Cond}_3$) and~($\mathrm{Cond}_4$). 
All the positions $Int_j$ are split into two classes:  the set $I_j'$ and its complement $Int_j \setminus I_j'$. For each position in $I'_j$ Charlie knows the corresponding bit from $X'$ sent by Alice. For the other bits Carlie already  knows the bits of $Y'_j$, but not the bits of $X'_j$. To  obtain $Z$, we should  invert at most $(\alpha + \delta) \cdot  |Int_j\setminus I'_j|$ bits of $Y'_{Int_j\setminus I'_j}$.
The number of such candidates is equal to 
\begin{align*}
S_A := \sum\limits_{s=0 }^{(\alpha + \delta) | Int_j\setminus I'_j | }
 {{| Int_j \setminus I'_j |} \choose {s}} \cdot 2^{ |Int_j \setminus I''_j|} 
= 2^{h(\alpha) (1-\lambda)  h(\alpha) k + O(\delta) k + O(\log k)}. 
\end{align*}
We set  the length of the hash function $\hash^A_l$ to 
$$ L_A = \log S_A + r  = (1-\lambda) h(\alpha) \lambda k + \kappa_1 \delta \cdot k + \kappa_2\log k+r$$
From Proposition~\ref{proposition-hash} it follows that the probability of the 2nd type error  $p_3^A\le 1/2^r$.

\medskip
 
We say that block $X_j'$ is \emph{reconstructible}, if the errors of type $1$, $2$, and $3$  do not happen. For each block $X_j'$,  probability to be non-reconstructible is at most $p^A_1(j) +  p^A_2(j) + p^A_3(j)$. This sum is less than  some threshold $\mu_A=\mu_A(n)$, where $\mu_A(n)\to 0$.  For the chosen values of parameters we have $\mu_A(n) = 1/(\log n)^c$ for some $c>0$.

Since the random permutations in the construction are $(m^c \cdot k)$-wise almost independent and the indices of hash functions are $m^c$-independent,   Proposition~\ref{proposition-lln} and Lemma~\ref{lemma1} give
an upper bound for  the probability  that  the fraction of non-reconstructible blocks is greater than $ 3\mu_A$: we conclude that this
probability is not  greater than $O(2^{-m^c})$.
Thus, with probability  $1- O(2^{-m^c})$ Charlie decodes on this stage all blocks of $X'_j$ except for at most $3\mu_A\cdot m$ of them.

Step ($7_C$) is similar to step ($3_C$). At this step Charlie can reconstructs $X'$ if the number of  non-reconstructible blocks $X'_j$ (at the previous step) is less than   $3\mu_A\cdot m$ (this is the number of errors that can be corrected by the Reed-Solomon checksums). Hence, the probability of failure at this step is less than  $O(2^{-m^c})$.

Step ($8_C$): No errors at this step.

Thus, with probability $1-O(2^{-m^d})$ (for some  $d<c$) Charlie successfully  reconstructs strings $X$ and $Y$.
\end{IEEEproof}

\section{Conclusion}

\emph{Possible applications.} We believe that the protocols suggested in the paper  (or their extensions) can be employed in  combinatorial versions of 
the \emph{omniscience} problem and the problem of multiparty  \emph{secret key agreement}.

\smallskip

\emph{Practical implementation.} The coding and decoding procedures in the protocol from  Theorem~\ref{thm-main} run in polynomial time. However, in the present form this protocol is not quite suitable for practical applications (at least for reasonably small $n$). Its most  time-consuming part is the use of the KNR generator from Proposition~\ref{proposition-knr}, which involves rather sophisticated computations. A simpler and more practical protocol can be implemented if we substitute $t$-wise almost independent permutations (the KNR generator) by  $2$-independent permutation (e.g., a random affine mapping). The price that we pay for this simplification is  only a  weaker bound for the probability of error, since with $2$-independent permutations we
have to employ Chebyshev's inequality instead of stronger versions of the law of large numbers (applicable to $n^c$-wise almost independent series of random variables).
(A similar technique was used in \cite{chuklin} to simplify the protocol from \cite{smith}.) In the simplified version of the protocol, the probability of error $\varepsilon(n)$ tends to $0$, but not exponentially fast.

\smallskip

\emph{Convergence to the optimal rate.} 
We have proved that the  communication complexity of our protocol is asymptotically optimal,  though
the convergence to optimal rate is rather slow (the $o(\cdot)$-terms in Theorem~\ref{thm-main} are not so small). 
This is a general weakness of the concatenated codes  and allied techniques, and there is probably no simple way to remedy it.

\smallskip

\emph{Open problem}:   Characterize  the set of all achievable pairs of rates for deterministic communication protocols.

\appendix

For the convenience of the readers 
and for the sake of self-containment, we proof in Appendix two results  that appeared implicitly in preceding papers,
see the discussion in Section~\ref{sec-2}.

\bigskip
\begin{IEEEproof}[Proof of Proposition~\ref{prop-chumbalov}]
To prove the proposition, it is enough to show that for every $\lambda\in(0,1)$ there exists a linear encoding scheme with messages of length
 $$
 \begin{array}{rcl}
 m_A &=& (\lambda+ h(2\alpha))n + o(n),\\
 m_B &=& (1-\lambda+ h(2\alpha))n + o(n).
 \end{array}
 $$
We use the idea of syndrome encoding that goes back to \cite{wyner}. First of all, we fix linear code with codewords of length $n$ that achieves the Gilbert--Varshamov bound. Denote $H=(h_{ij})$ ($i=1\ldots n$, $j=1\ldots k$ for $k\le h(2\alpha)n+o(n)$) the parity-check matrix of this code. So, the set of codewords $Z=z_1\ldots z_n$ of this code consists of all solutions of the system if uniform equations
 $$
 \left\{
 \begin{array}{rcl}
   h_{11} z_1 + h_{12}z_2 +\ldots +h_{1n} z_n &=&0,\\
    h_{21} z_1 + h_{22}z_2 +\ldots +h_{2n} z_n &=&0,\\
    \ldots\\
    h_{k1} z_1 + h_{k2}z_2 +\ldots +h_{kn} z_n &=&0\\
 \end{array}
 \right.
 $$
(a linear system over the field of $2$ elements). 
W.l.o.g. we assume that the rank of this system is equal to $k$, and the last $k$ columns of $H$ make up a minor of maximal rank (if this is not the case, we can eliminate the  redundant rows and re-numerates the columns of the matrix). 

\emph{Remark:} The assumption above guarantees that for every sequence of binary values $h^0_1,\ldots h^0_k$ and for any $z_1,\ldots, z_{n-k}$ we can uniquely determine  $z_{n-k+1},\ldots,z_n$ satisfying the linear system
 $$
 \left\{
 \begin{array}{rcl}
   h_{11} z_1 + h_{12}z_2 +\ldots +h_{1n} z_n &=&h^0_1,\\
    h_{21} z_1 + h_{22}z_2 +\ldots +h_{2n} z_n &=&h^0_2,\\
    \ldots\\
    h_{k1} z_1 + h_{k2}z_2 +\ldots +h_{kn} z_n &=&h^0_k\\
 \end{array}
 \right.
 $$
(in other words, $z_1,\ldots, z_{n-k}$ are the free variables of this linear system, and $z_{n-k+1},\ldots, z_n$ are the dependent variables).

Now we are ready describe the protocol. Denote $s=\lfloor \lambda \cdot (n-k) \rfloor$.

\textbf{Alice's message:} given $X=x_1\ldots x_n$, send to Charlie the bits $x_1, x_2,\ldots,x_{s}$ and the syndrome of $X$, i.e., the product $H \cdot X^\bot$.

\textbf{Bob's message:} given $Y=y_1\ldots y_n$, send to Charlie the bits $y_{s+1}, y_{s+2},\ldots,y_{n-k}$ and the syndrome of $Y$, i.e., the product $H \cdot Y^\bot$.

Let us show that Charlie can reconstruct $X$ and $Y$ given these two messages. First of all, given the syndromes $H \cdot X^\bot$ and $H \cdot Y^\bot$, Charlie obtains $H \cdot (X+Y)^\bot$, which is the syndrome of the bitwise sum of $X$ and $Y$.  Since the distance between $X$ and $Y$ is not greater than $\alpha n$, the bitwise sum $X+Y$ contains at most $\alpha n$ ones. The  code defined by matrix $H$ corrects $\alpha n $ errors, so Charlie can reconstruct all bits of $X+Y$ given the syndrome $H \cdot (X+Y)^\bot$.

Now Charlie knows the positions $i$ where $x_i \not= y_i$, and the bits $x_1 x_2 \ldots x_{s}$, and $y_{s+1} y_{s+2} \ldots y_{n-k}$. This information is enough to reconstruct the first $n-k$ bits in both strings $X$ and $Y$. Further, given the first $(n-k)$ bits of $X$ and $Y$ and the syndromes of these strings, Charlie  reconstructs the remaining bits of $X$ and $Y$ (see the remark above).
\end{IEEEproof}

\bigskip

The following proposition was implicitly proven in  \cite{orlitsky}. 
\begin{proposition}\label{prop-orlitsky}
For every small enough  $\alpha>0$   there exists a $\delta> 0$ such that for all large enough $n$,  
for the deterministic combinatorial Slepian--Wolf schemes with parameters $(\alpha,n)$
there is no achievable pairs of rates $(m_A,m_B)$ in the $(\delta n)$-neighborhoods of the points 
$(n, h(\alpha)n)$ and $(h(\alpha)n, n)$ (points $P_A$ and $P_B$ in Fig.~\ref{fig-4}).
\end{proposition}

\begin{IEEEproof}
At first we  remind the argument from  \cite[theorem~2]{orlitsky}. It  concerns an asymmetric version of the Slepian--Wolf scheme and it proves a lower bound for  the length of $Code_A(X)$ assuming that $Code_B(Y)=Y$. Here is the idea of the proof: for each value $c_A$, the set of pre-images $Code^{-1}_A(c_A)$ is  a set of strings with pairwise distances greater than $2\alpha n$, i.e., these pre-images make up an error correcting code that corrects $\alpha n$ errors. So we can  borrow from coding theory a suitable bound for the binary codes  and use it to bound the number of pre-images of $c_A$. Then, we obtain a lower bound for the number of values of  $Code_A(X)$ and, accordingly,  for the length of $Code_A(X)$. 
Technically, if we know from coding theory that for a binary code that corrects $\alpha n$ errors the number of codewords cannot be greater than 
 $
 (1-F(\alpha))n
 $
for some specific function $F(\alpha)$, then this argument implies that the length of  $Code_A(X)$ cannot be less than $F(\alpha) n$. For example, 
in the well known Elias--Bassalygo bound 
 $
  F(\alpha) = h(\frac12 - \frac12\sqrt{1-2\alpha}),
 $
 which is stronger than the trivial volume bound  $(1-h(\alpha))n$, \cite{bassalygo}. Alternatively, we can take the  McEliece--Rodemich--Rumsey--Welch bound.

Though this argument deals with only very special type of schemes where $Code_B(Y)=Y$, it  also implies some bound for the general Slepian--Wolf problem. Indeed, assume there exists a deterministic Slepian--Wolf scheme for string $X$ and $Y$ of length $n$ with $\dist(X,Y)\le T$ for some threshold $T=\alpha n$. Denote the lengths of Alice's and Bob's messages by 
 $$
 \begin{array}{rcl}
  m_A  &=&  (h(\alpha)+\delta_2)n,\\
  m_B  &=&  (1-\delta_1)n
  \end{array} 
 $$
respectively. We will prove that the pair of parameters $(\delta_1,\delta_2)$ cannot be too close to zero.
Notice that strings $X$ and $Y$ of any length $n'<n$ can be padded (by a prefix of zeros) to the length $n$. Hence,  the given communication scheme  (originally used for pairs of strings of length $n$, with the Hamming distance $T$) can be used also for the Slepian--Wolf problem with shorter strings  of length $n' = (1-\delta_1)n$ and  the same distance between words  $T$  (which can be represented as $T=\alpha'n$ for $\alpha'=\frac{\alpha}{(1-\delta_1)}$). Thus, for the Slepian--Wolf problem with parameters $(n',\alpha')$ we have a communication scheme with messages of the same lengths $m_A$ and $m_B$, which can be represented now as  
 $$
 \begin{array}{rcl}
   m_A  &=&  \frac{h(\alpha)+\delta_2}{1-\delta_1}n',\\
   m_B &=&  n'.
  \end{array} 
 $$
We apply the explained above Orlitsky--Viswanathan bound to this  scheme  and obtain
 $$
    \frac{h(\alpha)+\delta_2}{1-\delta_1}  \ge F\left( \frac{\alpha}{(1-\delta_1)} \right)
 $$
(for any suitable bound $F(x)$ from coding theory). 
It follows that 
 \begin{eqnarray}
   \delta_2  \ge (1-\delta_1 )F\left( \frac{\alpha}{(1-\delta_1)} \right)- h(\alpha).\label{eq-orlitsky}
 \end{eqnarray}
The functions $F(x)$ from the Elias--Bassalygo bound and from the McEliece--Rodemich--Rumsey--Welch bound
are a continuous functions, and for small positive $x$ they are bigger than $h(x)$.
Hence,  (\ref{eq-orlitsky}) implies that for every fixed $\alpha$ the values of $\delta_1$ and $\delta_2$ cannot be very small simultaneously.

We do not discuss here the exact shape of this forbidden zone for values of $(\delta_1,\delta_2)$;  we only conclude that  small  neighborhoods around the point $(h(\alpha)n, n)$ and (from a symmetric argument) $(n, h(\alpha)n)$ cannot be achieved,
which concludes the proof of the proposition.
\end{IEEEproof}
\newpage

\end{document}